\documentclass[sigconf]{acmart}

\usepackage{booktabs} 
\usepackage{graphicx}
\usepackage{wrapfig}
\usepackage{subcaption}
\usepackage{tabularx}
\usepackage{arydshln}
\usepackage{amsmath}
\usepackage{breakurl}
\usepackage{multirow}
\usepackage{amsmath}
\usepackage{amsfonts}
\usepackage{amssymb}
\usepackage[shortlabels]{enumitem}
\usepackage{balance}
\setcopyright{rightsretained}

\copyrightyear{2018} 
\acmYear{2018} 
\setcopyright{acmcopyright}
\acmConference[KDD '18]{The 24th ACM SIGKDD International Conference on Knowledge Discovery & Data Mining}{August 19--23, 2018}{London, United Kingdom}
\acmBooktitle{KDD '18: The 24th ACM SIGKDD International Conference on Knowledge Discovery \& Data Mining, August 19--23, 2018, London, United Kingdom}
\acmPrice{15.00}
\acmDOI{10.1145/3219819.3219863}
\acmISBN{978-1-4503-5552-0/18/08}
\author{Tara Safavi}
\affiliation{\institution{University of Michigan, Ann Arbor}}
\email{tsafavi@umich.edu}

\author{Maryam Davoodi}
\affiliation{\institution{Purdue University}}
\email{mdavoodi@purdue.edu}

\author{Danai Koutra}
\affiliation{\institution{University of Michigan, Ann Arbor}}
\email{dkoutra@umich.edu}

\begin{document}
\title{Career Transitions and Trajectories: A Case Study in Computing}
  
\newtheorem{observation}{Observation}
\newtheorem{problem}{Problem}
\newtheorem{algo}{Algorithm}

\newcommand{\hide}[1]{}
\newcommand{\notice}{\todo}
\newcommand{\reminder}[1]{{\textsf{\textcolor{red}{[#1]}}}}
\newcommand{\dk}[1]{{\textsf{\textcolor{blue}{DK: #1}}}}
\newcommand{\ts}[1]{{\textsf{\textcolor{orange}{[#1]}}}}
\newcommand{\nj}[1]{{\textsf{\textcolor{green}{[NJ: #1]}}}}
\newcommand{\new}[1]{{{\textcolor{blue}{#1}}}}
\newcommand{\vectornorm}[1]{\left|\left|#1\right|\right|}

\newcommand{\datasetSize}{6,781}
\newcommand{\modelName}{$\textbf{R}^{3}$}
\newcommand{\timestamp}[1]{t} % _{#1}}
\newcommand{\edge}[2]{(#1, #2, \timestamp{i})}
\newcommand{\employee}{p}
\newcommand{\flowVolume}[2]{W_f^{\timestamp{i}}({#1\rightarrow#2})}
\newcommand{\newFlowVolume}[2]{W_{\mathbf{R}^{3}}^{\timestamp{i}}({#1\rightarrow#2})}
\newcommand{\resourcesaverage}{\overline{R_{SRC}(u, v, \timestamp{i})}}
\newcommand{\resources}{R_{SRC}}
\newcommand{\retention}{R_{TN}}
\newcommand{\relativeGrowth}{R_{GR}}
\newcommand{\sector}[1]{\sigma(#1)}
\newcommand{\careerLength}[2]{\ell(#1, #2)}
\newcommand{\careerLengthAt}[3]{\ell(#1, #2, #3)}
\newcommand{\experience}{\textrm{experience}}
\newcommand{\netHires}{\textrm{net\_hires}}
\newcommand{\transitionNetwork}{\mathcal{G}_{f}}
\newcommand{\nodes}{\mathcal{V}}
\newcommand{\edges}{\mathcal{E}}
\newcommand{\nodeA}{\textsc{STABLE-LLC}}
\newcommand{\nodeB}{\textsc{UNI}}
\newcommand{\nodeC}{\textsc{DECLINE-LLC}}
\newcommand{\nodeD}{\textsc{STARTUP}}
\newcommand{\egoFeatures}{\textbf{IND}}
\newcommand{\orgFeatures}{\textbf{ORG}}
\newcommand{\sysFeatures}{\textbf{\modelName{}}}

\newcommand{\mkclean}{
    \renewcommand{\reminder}{\hide}
    \renewcommand{\todo}{\hide}
    % \usepackage[top=1in, bottom=1in, outer=0in, inner=0in, heightrounded, margin
    % parwidth=0in, marginparsep=0in]{geometry}
    % \usepackage[letterpaper]{geometry}
    % \PassOptionsToPackage{letterpaper}{geometry}
    \setlength{\marginparwidth}{0in}
    \setlength{\marginparsep}{0in}
}

% from cf: shorthands - also they make tighter lists
% \newcommand{\bit}{\begin{itemize*}}
% \newcommand{\eit}{\end{itemize*}}
% \newcommand{\ben}{\begin{enumerate*}}
% \newcommand{\een}{\end{enumerate*}}
\newcommand{\bit}{\begin{compactitem}}
\newcommand{\eit}{\end{compactitem}}
\newcommand{\ben}{\begin{compactenum}}
\newcommand{\een}{\end{compactenum}}

\newcommand{\QED}{ \hfill {\bf QED}}
\newcommand{\method}{\textsc{OurMETHOD}\xspace}

\newcommand{\placeholder}{{............................................................................................................\\
............................................................................................................\\
............................................................................................................\\
............................................................................................................\\
............................................................................................................\\
............................................................................................................\\
............................................................................................................\\
............................................................................................................\\}}

\begin{abstract}
From artificial intelligence to network security to hardware design, it is well-known that computing research drives many important technological and societal advancements.
However, less is known about the long-term career paths of the people behind these innovations.
What do their careers reveal about the evolution of computing research?
Which institutions were and are the most important in this field, and for what reasons?
Can insights into computing career trajectories help predict employer retention?

In this paper we analyze several decades of post-PhD computing careers using a large new dataset rich with professional information, and
propose a versatile career network model, \modelName{}, that captures temporal career dynamics.
With \modelName{} we track important organizations in computing research history, analyze career movement between industry, academia, and government, and build a powerful predictive model for individual career transitions.
Our study, the first of its kind, is a starting point for understanding computing research careers, and may inform employer recruitment and retention mechanisms 
at a time when the demand for specialized computational expertise far exceeds supply.

\end{abstract}

% \settopmatter{printfolios=true} % comment out to drop the page numbers
\maketitle

\section{Introduction}
\label{sec:introduction}
% Important source to incorporate: https://computinged.wordpress.com/2017/12/13/resources-for-dealing-with-the-undergraduate-cs-capacity-crisis-guest-post-from-eric-roberts/

From the invention of the Unix operating system in the 1970s to the ongoing artificial intelligence revolution, the importance and impact of computing research can hardly be overstated.
The world has taken notice accordingly: the news media regularly covers everything from frontiers in computer design~\cite{economistmoores} to the earnings of AI experts~\cite{nytimesai}.
Naturally, questions regarding computing research careers are becoming relevant.
What happens after a PhD in computer science?
Which organizations are, or were, central in computing research?
How do expertise and talent flow between organizations?

In this study, we answer these questions by analyzing a unique career trajectory dataset of computer science PhD graduates from the 1970s to the present.
Our goal, broadly, is to understand the evolution of computing research as a profession on the levels of individual \textbf{career transitions} (movement between distinct employers),
\textbf{organizations} (employers), and three respective \textbf{sectors} (industry, academia, and government).
To do so we propose \modelName{}, a versatile career network model that captures \emph{resource} flow, employer \emph{retention}, and \emph{relative} organizational growth. %  in the context of career transitions.
Combining \modelName{} with the HITS link analysis algorithm~\cite{kleinberg1999hits}, which has not (to the best of our knowledge) been used in career analysis before, we demonstrate \modelName{}'s versatility with insights of varying granularity:

\setlist[itemize]{leftmargin=*}
\begin{itemize}
    \item \textbf{System-wide evolution}. We identify key organizations, from startups to universities to industry leaders, in computing research history.
    \modelName{} captures crucial factors beyond size and popularity that contribute to organizational ``importance'', demonstrating that \emph{some organizations are important precisely for their small sizes, low retention, or short existences}.
    \item \textbf{Cross-sector career movement}. We examine post-PhD career transitions across sectors.
    Beyond finding evidence that cross-sector collaboration is increasing, we use \modelName{} to reveal \emph{significant asymmetry in the frequency, timing, and ``prestige'' of career moves between academia and industry}.
    \item \textbf{Individual retention prediction}. Finally, we predict career transitions by combining \modelName{} network dynamics and individual career trajectory information.
    We demonstrate \modelName{}'s immediate utility in boosting prediction power with \emph{interpretable features that can inform employer recruitment and retention mechanisms}.
\end{itemize}

This work is a starting point for large-scale studies of computing career trajectories.
Such analyses are becoming crucial as demand for computing expertise grows and our world increasingly depends on research innovations in computer science.

\vspace{.1cm}
\noindent \textbf{Outline}.
This paper is organized as follows: we first discuss some of our extensive data standardization pipeline and describe our post-processed dataset (Sec.~\ref{sec:data}).
We then motivate and detail our \modelName{} career network model (Sec.~\ref{sec:model}).
With  \modelName{} we analyze computing research careers at several levels of granularity (Sec.~\ref{sec:transitions-trajectories}).
Finally, we outline related areas of work and discuss future directions based on our study's results and limitations (Secs.~\ref{sec:related} through \ref{sec:conclusion}).

\section{Data}
\label{sec:data}
\begin{figure*}
	\begin{subfigure}{.48\textwidth}
		\centering
		\includegraphics[width=.9\linewidth]{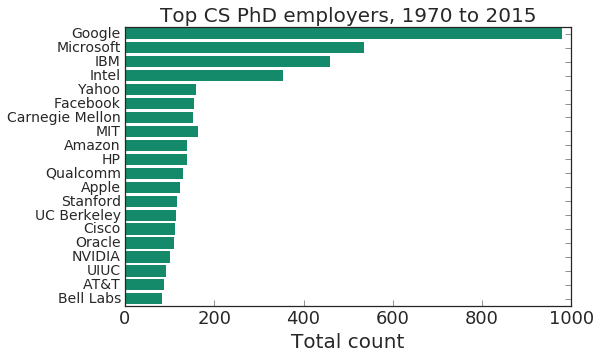}
		\caption{Top post-PhD employers.}
		\label{fig:all-employers}
	\end{subfigure}
	\begin{subfigure}{.46\textwidth}
		\centering
        \includegraphics[width=.89\linewidth]{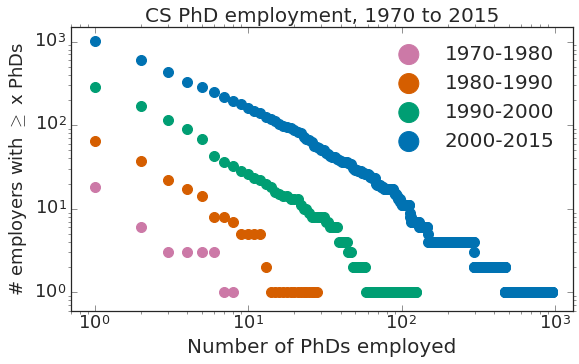}
        \caption{Pareto cumulative distribution of PhD employment.}
		\label{fig:employment-powerlaw}
	\end{subfigure} 
	\vspace{-.2cm}
	\caption{Most CS PhDs concentrate at one of a few institutions: post-PhD employment distribution follows a power law.}
	\vspace{-.2cm}
	\label{fig:phd-employment}
\end{figure*}

\begin{figure}[t]
	\centering
	\vspace{-.2cm}
	\includegraphics[width=.95\linewidth]{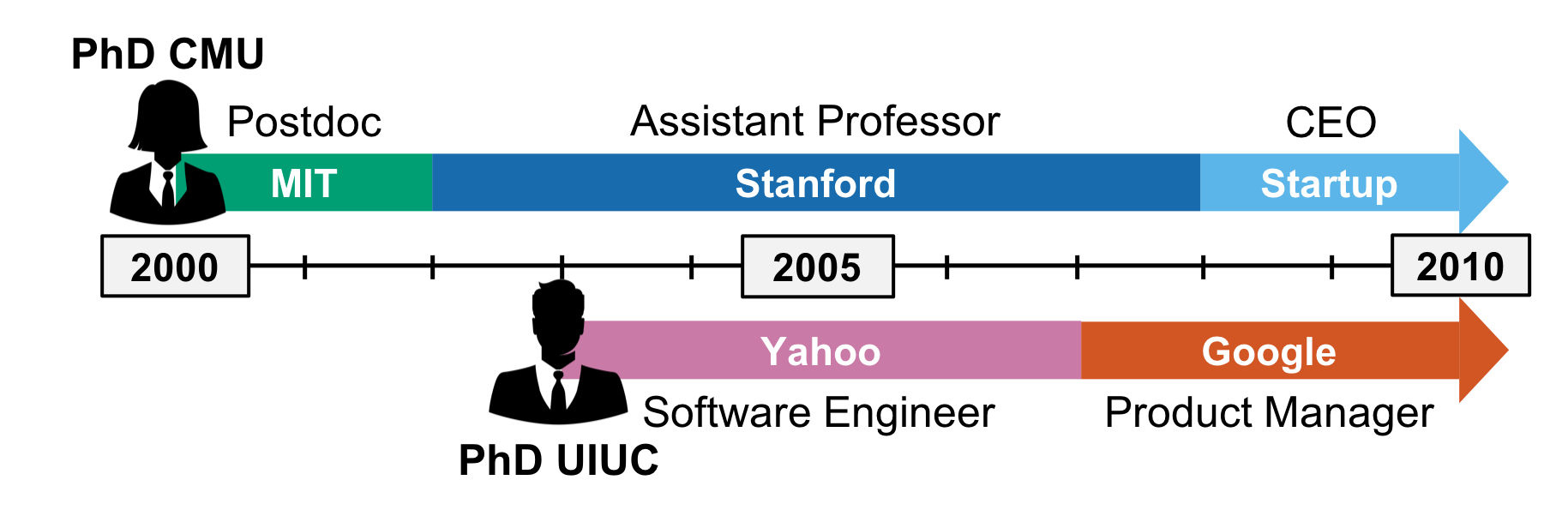}
	\vspace{-.3cm}
	\caption{Fictitious but plausible samples from our dataset.}
% 	\vspace{-.4cm}
	\label{fig:fake-data}
\end{figure}

\vspace{.1cm}
\noindent \textbf{Data collection}. 
To obtain our data, we automatically crawled the public online information of around 10 thousand PhD graduates from the 1970s to 2015 in computer science and related subfields.
We matched these graduates from the Proquest Digital Library of PhD dissertations to an online public professional (LinkedIn) profile.
% The PhD graduation years in our dataset span 1964 to 2015, inclusive.
To guide automatic data collection, % and obtain necessary knowledge about which computer science programs graduate the most PhDs, 
we obtained data for those with PhDs
from the top 50 US computer science graduate programs as specified in the 2014 \emph{US News \& World Report} (USNWR)\footnote{\url{https://www.usnews.com/best-graduate-schools/top-science-schools/computer-science-rankings}}.
We do not use the actual USNWR \emph{rankings}, which have been criticized~\cite{bastedo2010usnews}, anywhere in our study.
% For privacy, our dataset is anonymized.
Per person, we retained the PhD school, graduation year, and all available post-PhD employers, job start/end dates, and job titles.
Figure~\ref{fig:fake-data} gives two illustrative samples.

While it is not possible to fully verify whether online professional profiles are up-to-date or truthful, we manually validated our data by inspecting individuals' listed employers.
We discarded profiles with suspicious employers: for example, overly generic names like ``college'' or companies with no online records of existence.
Moreover, because we collected the profiles of people whose dissertations were verified by ProQuest, it is unlikely that their profiles were fake or set up by fake accounts.
After validation, we retained 17,358 unique employment records for \datasetSize{} PhDs over five decades.
We make two anonymized versions of the data available at~\url{https://github.com/tsafavi/career-transitions-data}.

\vspace{.1cm}
\noindent \textbf{Name and sector standardization}. Using several publicly available collections of organization names\footnote{\url{https://www.crunchbase.com}}$^{,}$\footnote{\url{http://www.nasdaq.com/screening/company-list.aspx}}, we created a centralized list of academia, industry, and government organizations.
We standardized each organization in our dataset % using substring matching
against this list.
For instance, all variants of Microsoft---i.e., Microsoft Bing, Microsoft Skype, and Microsoft Research---were grouped under one umbrella.
% We also collected job titles to retain information about research-related roles, like research scientist positions, in all three sectors. % \ts{Will this data be used?}
% There is no centralized public database of company mergers, acquisitions, or spinoffs, so we let users specify their ``parent'' employer in these cases.
% For example, an employer specified as ``Altera, acquired by Intel'' became Intel;
%  by contrast, an employer listed as ``Altera'', which existed as a company for several decades before its acquisition by Intel, did not.
In the case of ambiguity, we automatically mapped names to their most well-known instances in computing, like ``CMU'' as an accepted acronym for Carnegie Mellon University. % rather than Central Michigan University.
Universities without specified campuses mapped to their flagships.
For example, the University of Michigan listed without one of Ann Arbor, Flint, or Dearborn became the University of Michigan Ann Arbor.

% \paragraph{Sector labeling}
% \vspace{.1cm}
% \noindent \textbf{Sector labeling}.
To categorize employers into sectors (one of industry, academia, or government), we used our centralized organization list, keywords like ``LLC'' and ``college'', and rule-based automatic classification.
For the 444 organization names that our system failed to categorize, we provided 6 expert assessors with those organizations and a set of publicly available rules\footnote{\url{http://bit.ly/2ErexBh}}.
% described in Table~\ref{table:sectors}.
The inter-rater agreement on a subset of those employer names was 73.6\% using Fleiss' kappa~\cite{fleiss1971kappa}, which quantifies the degree of inter-rater agreement over that expected by chance.
This ``substantial'' inter-rater agreement~\cite{landis1977agreement} demonstrates the relative simplicity and interpretability of our rule-based system.

% \subsection{Dataset description}
% \label{sec:dataset-description}
\begin{figure}% [h!]
	\centering
	\includegraphics[width=.99\linewidth]{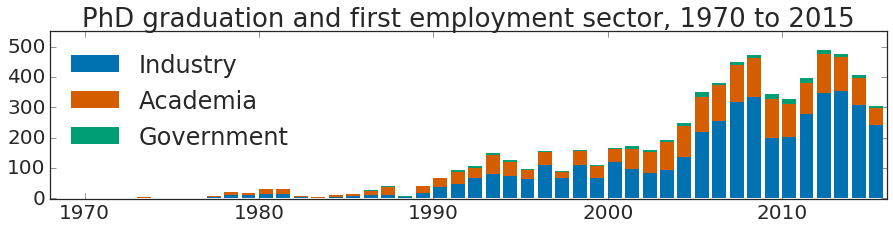}
% 	\vspace{-.2cm}
	\caption{First employment sector of PhDs per year.} %production and first employment.}
% 	\vspace{-.4cm}
	\label{fig:phd-grads-first-sector}
\end{figure}
\vspace{.1cm}
\noindent \textbf{Description}.
% \vspace{.1cm}
% \noindent 
\emph{(1) Education.}
% Our dataset comprises PhD graduates in electrical engineering and computer science.
The school with the largest graduate representation in our dataset, at 6.7\%, is the University of Illinois Urbana-Champaign (UIUC).
Carnegie Mellon, MIT, UC Berkeley, and Stanford follow closely. 
As expected, PhD production volume increased substantially in the last 15 years (Fig.~\ref{fig:phd-grads-first-sector}), during which interest in computing became widespread and compute resources and data availability skyrocketed.
% Nearly three-fourths of the graduates in our dataset obtained a PhD after 2000.
The rate of computer science PhD graduation between 2003 and 2008 grew on average $18.33\%$ per year until a peak volume in 2008, then dropped briefly around the beginning of the Great Recession.
Of these graduates, 11.7\% had postdoctoral experience, with an average of 1.13 postdocs per person.
The volume of PhD graduates beginning postdocs spiked between 2009 and 2012, again with the onset of the recession.

While representation bias in our dataset is possible, as collecting data on \emph{all} graduating CS PhDs is difficult, our data are corroborated by the Computing Research Association's Taulbee Survey\footnote{\url{https://cra.org/resources/taulbee-survey/}}.
This survey collects, among other figures, data from US and Canada higher education institutions on graduating PhDs in computer science and related fields.
The trend of our PhD production data strongly correlates (Pearson $r=0.75$) with the Taulbee PhD production numbers publicly available from 2002 to 2015.
% As we only include 50 universities out of 205 listed by the CRA, the number of graduates in our dataset from 2002 to 2015 is 25.2\% that of the corresponding Taulbee total.

\vspace{.1cm}
\noindent \emph{(2) Employment.}
The sector distribution of organizations in our dataset is 83.5\% industry, 14\% academia, and 2.5\% government. 
Google is by far the most popular employer (Fig.~\ref{fig:all-employers}), with nearly 15\% of the entire dataset having worked there at least once since its inception in the late 1990s.
The most popular destination in academia, at around $1\%$ of all PhDs in the dataset, is Carnegie Mellon.
Like many other well-documented phenomena~\cite{FaloutsosFF99,ClausetSN09}, computer science PhD employment among organizations appears to follow a power-law distribution (Fig.~\ref{fig:employment-powerlaw}), demonstrating that most computing PhD talent has concentrated in a few companies and universities.
% \reminder{The distributions of PhD employment for different timeframes are depicted in Fig.~\ref{fig:employment-powerlaw}}.
% ~\ts{compute slope?}

Although a PhD is often considered a gateway to academia, a majority of computer science PhDs in our dataset immediately work in industry.
On average, 57\%  go to industry, 39\% go to academia, and 4\% go to government per year (Fig.~\ref{fig:phd-grads-first-sector}).
However, while industry jobs are more popular, academic jobs have higher longevity.
The mean retention rate for industry employers in our dataset is 4.65 years; for academia, 5.84 years; for government, 4.91 years, with significant differences between academia and the others ($p\ll.00001$ academia/industry, and $p=0.002$ %2\times10^{-3}$ 
academia/government, two-sided $t$-test).
While this may be related in part to academic tenure policies, we do consider postdoctoral positions at academic institutions, which are intended to be short, and positions beyond tenured professorships as part of academia here.
% Furthermore, nearly 40\% of all job titles for non-academia jobs are related to research or higher education, demonstrating the abundance and diversity of cross-sector opportunities available after a computing PhD.

\section{\modelName{} transition network model}
\label{sec:model}
To analyze the evolution of computing research with our unique dataset, we need an employer ``desirability'' or ``importance'' measure for computing PhDs.
Such a measure quantifies hierarchies between organizations and helps us anchor our analysis around key representative institutions of the profession.
For this two components are necessary: (1) a \emph{network representation} that captures the dynamics of career paths; 
and (2) an \emph{organizational ranking method} that captures both employee influx and outflux.

\vspace{.1cm}
\noindent \textbf{(1) Network representation.}
Among the various ways to model trajectory or sequence data~\cite{xu2016higherorder,KoutraBH15,SafaviSK17,BrugereGB18}, a natural first choice is the transition network, which is a directed graph that here captures the post-PhD career transitions between employers (states). 
This representation is often called an \textbf{aggregate flow network} or ``talent flow graph'' or ``job transition/hop network''~\cite{kapur2016unirankings, xu2016talentcircles, oentaryo2017jobhops}. 
In this representation $\transitionNetwork(\nodes, \edges)$, 
each node $v \in \nodes$ is an industry, academia, or government organization.
Each directed edge $\edge{u}{v} \in \edges$ is a set of  employee transitions from organization $u \in \nodes$ to organization $v \in \nodes$ in year $\timestamp{i}$.
The weight of edge $\edge{u}{v}$, which we denote as $\flowVolume{u}{v}$, captures the total number of employees making a career transition from $u$ to $v$ during year $\timestamp{i}$. 

The aggregate flow network $\transitionNetwork$ is simple and interpretable.
However, it can obscure important insights, which we demonstrate in Sec.~\ref{sec:transitions-trajectories}, for several reasons.
For one, our data show that most PhD-trained talent in computing concentrates in very few organizations (Fig.~\ref{fig:employment-powerlaw}).
Ranking organizations by aggregate flow \emph{heavily} favors these organizations, which are mostly large companies, whereas organizational size is but one determinant of importance in the real world.
Furthermore, capturing only aggregate transition volume cannot answer important temporal questions encoded in career sequence data.
Which organizations have higher turnover than normal?
Which are growing quickly relative to their size?
Which are desirable for fresh graduates versus senior engineers, distinguished researchers, and program directors?
To answer these questions, we propose the \modelName{} \textbf{transition network} model.
Each \emph{R} in \modelName{} transforms $\transitionNetwork$'s edge weights  to capture a specific career dynamic, which we define as \emph{resource} transfer ($\resources$), employee \emph{retention} ($\retention$), and \emph{relative} organizational growth ($\relativeGrowth$).

\begin{figure}[t!]
	\centering
	\includegraphics[width=.9\linewidth, trim={.1cm .1cm .1cm .1cm},clip]{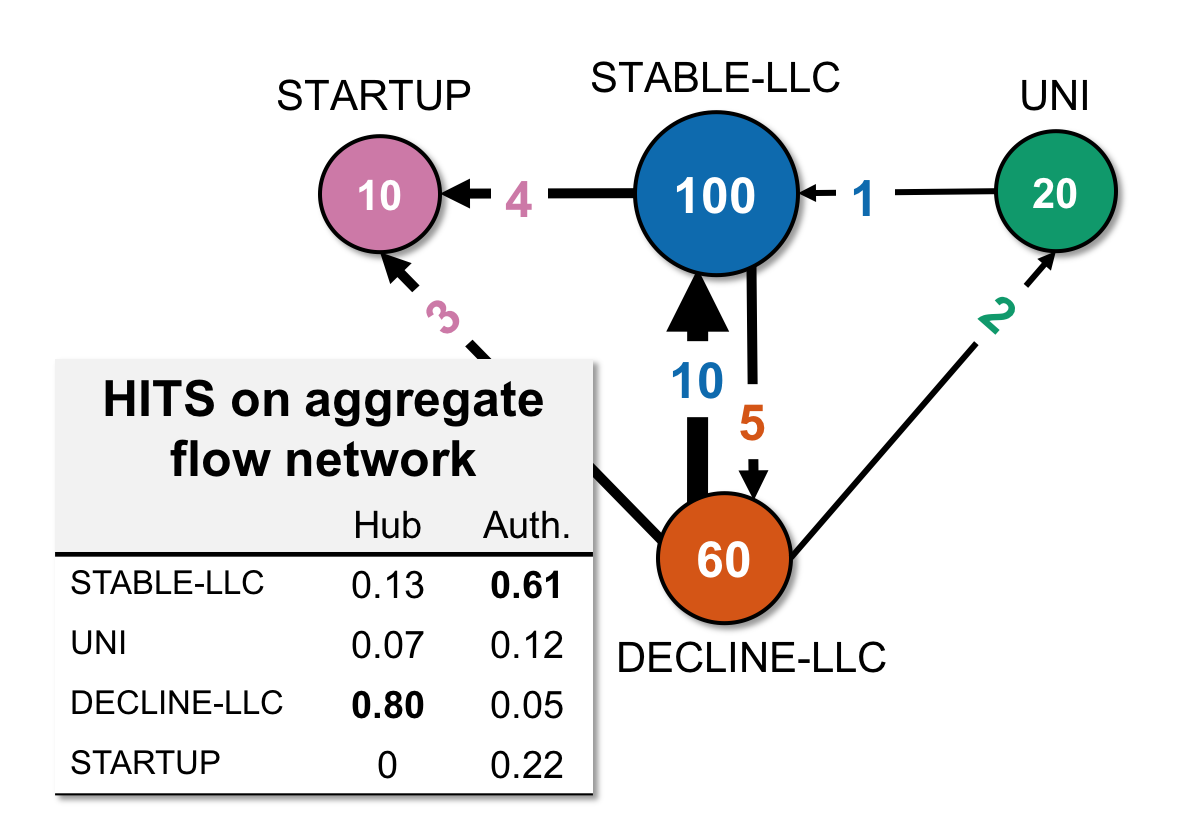}
	\caption{A hypothetical transition network comprising a stable company \nodeA{}, a university \nodeB{}, a declining company \nodeC{}, and a fast-growing \nodeD{}. Each node is labeled with the number of employed PhDs before transitions.
	The edge weights denote the number of PhDs moving to or from each node. 
	}
	\vspace{-.4cm}
	\label{fig:r3-model}
\end{figure}

\vspace{.1cm}
\noindent \textbf{(2) Ranking.} PageRank and HITS are two of the most well-known node centrality measures on directed graphs~\cite{getoor2005linkmining}.
The former, which has been used in career mining~\cite{kapur2016unirankings, oentaryo2017jobhops} and many other settings~\cite{TongFP06,Gleich15,KoutraDBWIFB17,KoutraF17}, ranks nodes by the quantity and ``importance'' of their in-links, and outputs one set of scores.
By contrast, HITS~\cite{kleinberg1999hits} outputs two sets of scores, one for \textbf{hubs} and one for \textbf{authorities}.
Hubness measures each node's ``indexing power'' by the number and strength of its outgoing links to authority nodes. 
Authority measures each node's ``relevance'' by the number and strength of its incoming links from hub nodes.
% Note that these definitions are inherently recursive, as is the PageRank formulation, but both are computed with iterative algorithms converge.

While the \modelName{} model proposed in this section can work with PageRank, HITS, and any other link analysis algorithm on directed weighted graphs, we design it with HITS in mind.
In career analysis, both the in- \emph{and} out-links of nodes, which respectively capture organizational influx and outflux, characterize employer roles and rankings in the flow graph.  
For this reason we posit that 
{\bf identifying both hubs and authorities
best captures the natural meaning of career transitions}, and the full dynamics of transitions from an organizational perspective.
Intuitively, authority organizations attract talent and expertise from hub organizations.
% In our narrative, we thus treat ``hubs''/```producers'', and ``authorities''/``consumers'', as synonyms.
For brevity, we do not cover the theory of HITS (see~\cite{kleinberg1999hits} for details).

% \vspace{0.15cm}
In the remainder of this section we define \modelName{} and demonstrate its effects on a simple but plausible example (Fig.~\ref{fig:r3-model}).
As job transition frequencies have been shown to follow a yearly cyclic pattern~\cite{xu2016talentcircles}, from here on we assume yearly time units. 
% In the next section, we show how applying the \modelName{} transformations to $\transitionNetwork$ leads to insights into the evolution of computing research.
% For reproducibility and further analysis, we will make the model and experimental code available upon acceptance.

\subsection{$\resources$: Modeling resources} 
\label{sec:resources}

% \ts{Say something about how an organization's resources can be defined by the strength or skill of their employees}
Our first feature, \textbf{resources} ($\resources$), captures \textbf{the level of cumulative employee expertise} in inter-organization transitions.
The intuition of $\resources$ is that a longer career leads to more advanced individual expertise and organizational value.
For example, one might rise from a software engineer to a directorate role, or else from assistant to full professorship, with time. 
Our goal with $\resources$ in terms of HITS is to capture organizational hubs and authorities for experienced people, who are skilled \emph{resources}.

\begin{figure}[t]
	\centering
	\includegraphics[width=.7\linewidth]{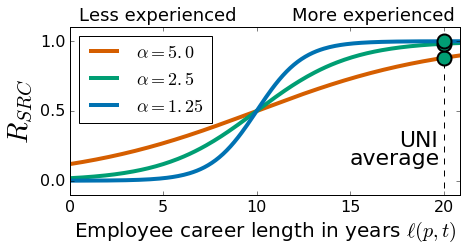}
	\caption{Computing $\resources$ for \nodeB{}'s transitions (Fig.~\ref{fig:r3-model}), assuming a system-wide average career length of 10 years.}
 	\vspace{-.4cm}
	\label{fig:sigmoid-curve}
\end{figure}

To quantify each employee's expertise level, we use a variant of the logistic skill-gain model from economics and organizational theory~\cite{yelle1979learningcurve}. 
In more detail, as shown in Fig.~\ref{fig:sigmoid-curve}, we model the expertise level of a PhD $p$ making a career transition in year $\timestamp{i}$  as a sigmoid function of her career length up to that year, $\careerLength{\employee}{\timestamp{i}}$: 

\begin{align*}
    \resources(\employee, \timestamp{i}) &= \left(1 + \textrm{exp}[{-\frac{\careerLength{\employee}{\timestamp{i}} - \overline{\ell(\timestamp{i})}}{\alpha}}]\right)^{-1}. 
\label{eq:resources}
\end{align*}

\noindent 
In the above formulation, $\alpha$ controls the curve's steepness and $\overline{\ell(\timestamp{i})}$, the sigmoid midpoint, is the system-wide average career length at year $\timestamp{i}$ (10 years in Fig.~\ref{fig:sigmoid-curve}).
$\resources$ thus scores each transitioning PhD based on her experience relative to her peers.
The transitions of those who entered the system earlier are deemed more valuable for the source and target organizations, although in our examples and analyses
we set $\alpha=\overline{\ell(\timestamp{i})}/{2}$ (the least steep curve in Fig.~\ref{fig:sigmoid-curve}, orange) to avoid over-penalizing those with fewer years of experience. 

We transform each directed edge of the aggregate flow network $\transitionNetwork$ to concentrate flow in the graph around the movement of experienced people:
\begin{align}
    \newFlowVolume{u}{v}&=\flowVolume{u}{v} \cdot \resourcesaverage \\ \nonumber
    &= \sum\limits_{\textrm{PhD } p: \hspace{1mm} u \rightarrow v | t} \resources(p, t),    
\end{align}
where $\resourcesaverage$ is the average $\resources$ score of employees $p$ moving from the source node $u$ to the target node $v$ during year $\timestamp{i}$, each denoted as $\textrm{PhD } p: \hspace{1mm} u \rightarrow v | t$ above. 
Note  that our logistic model does not account for skill loss over time.
While ``productivity decline'' in academia has been studied for research publication rates over time, among other phenomena, this narrative has been recently questioned~\cite{Way2016narrative}. 
As such we do not include it in our model.

\begingroup
\setlength{\intextsep}{5pt}
\setlength{\columnsep}{7pt}%
\begin{wrapfigure}{r}{0.18\textwidth}
    \vspace{-.1cm}
	\centering
	\includegraphics[width=0.92\linewidth, ,trim={0cm 0cm .05cm 0cm},clip]{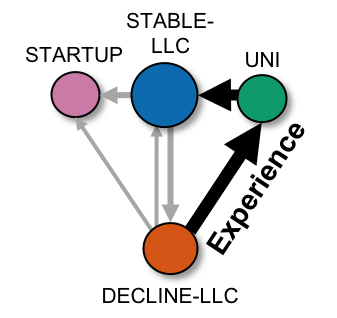}
	\vspace{-10pt}
	\caption{$\resources$ effects.}
% 	\vspace{-.2cm}
	\label{fig:resources}
\end{wrapfigure}
\vspace{.1cm}
\noindent \textbf{Example.}
Assume that the few people transitioning to and from \nodeB{} (green, Figs.~\ref{fig:r3-model} and~\ref{fig:resources}) are distinguished professors and industry leaders with 20 years of experience.
Also, assume all others among \nodeA{}, \nodeC{}, and \nodeD{} have 5 years of experience.
Given a system-wide average career length of 10 years, transforming $\transitionNetwork$ with $\resources$ results in \nodeB{}'s % $\resources$
authority score increasing from 0.12 to 0.31, reflecting its centrality as an employer of highly skilled people.

\endgroup

\subsection{$\retention$: Modeling retention}
\label{sec:retention}

\textbf{Retention} ($\retention$) captures how well \textbf{organizations retain talent}, which has been shown to be crucial in career transition graphs~\cite{kapur2016unirankings}.
Indeed, inter-employer transitions alone are comparatively sparse, with only 22\% of our dataset transitioning on average per year.
Our motivation for $\retention$ is that low retention may signify a variety of real-world meanings in organizations, from undesirability to a short existence (i.e., startups that fail or get acquired quickly).
In the context of HITS, we use $\retention$ to identify low-retention hub organizations that serve as ``stepping stones'' to other authorities.

Since employers with higher retention are better able to develop their employees' job-specific skills, we model retention by capturing  organizational ``expertise'' on a sigmoid curve.
To first account for significant differences in sector retention rates (Sec.~\ref{sec:data}), we stratify organizations by sector.
We then model the retention of an organization $v$ at year $\timestamp{i}$ as a sigmoid function, comparing $v$'s average PhD retention rate at year $\timestamp{i}$, $\overline{\ell(v, \timestamp{i})}$, to its \emph{sector}'s
current average PhD retention rate $\overline{\ell(\sector{v}, \timestamp{i})}$, where $\sector{v}$ is $v$'s sector:
\begin{align}
    \retention(v, \timestamp{i}) &= \left(1 + \textrm{exp}[{-\frac{\overline{\ell(v, \timestamp{i})} - \overline{\ell(\sector{v}, \timestamp{i})}}{\beta}}]\right)^{-1}. \nonumber
\label{eq:retention}
\end{align}
The idea here is that employers with higher-than-average PhD retention in their sector receive a higher $\retention$ score, and vice versa, although as before we smooth the curve by setting $\beta=\overline{\ell(\sector{v},\timestamp{i})}/2$.
Our goal with $\retention$ is to capture hubs, so we transform \emph{outgoing} edges for each node $v$ in $\transitionNetwork$ with $1 - \retention(v, \timestamp{i})$:
\begin{align}
    \newFlowVolume{v}{u}=\flowVolume{v}{u} \cdot (1 - \retention(v, \timestamp{i})).
\end{align}
Taking the converse increases outflux from low-retention employers and decreases outflux from high-retention employers.
% This in effect ``increases traffic'' from the former and ``blocks traffic'' toward the latter.

\begingroup
\setlength{\intextsep}{0pt}
\setlength{\columnsep}{7pt}%
\begin{wrapfigure}{r}{0.18\textwidth}
	\centering
	\includegraphics[width=.79\linewidth,trim={0cm 0.4cm .05cm 0cm},clip]{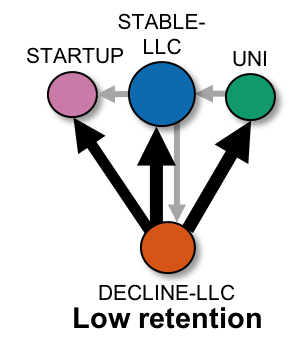}
	\vspace{-.2cm}
	\caption{$\retention$ effects.}
	% \vspace{-.5cm}
	\label{fig:retention}
\end{wrapfigure}
\vspace{.1cm}
\noindent \textbf{Example.}
Returning to Fig.~\ref{fig:r3-model}, assume that \nodeA{}'s average retention matches the industry average; \nodeC{}'s is around one-half the industry average due to its decline; and \nodeB{}'s is twice the academia average due to its prestige and tenure policies. 
By transforming the flow network $\transitionNetwork$ with  $1 - \retention$, \nodeC{}'s hub score increases to 0.91 and  \nodeB{}'s hub score drops to 0.01, magnifying the respective retention abilities of these institutions.

\endgroup

\subsection{$\relativeGrowth$: Modeling relative growth} 
\label{sec:relative-growth}

Our last feature, \textbf{relative growth} ($\relativeGrowth$), quantifies \textbf{growth relative to organization size}.
The goal of $\relativeGrowth$ is to boost the authority of small, fast-growing organizations.
Its HITS interpretation is that employers with high $\relativeGrowth$, like buzzworthy startups or fast-growing university computer science departments, should gain authority even with relatively low influx.

Extending the literature in ecology and stock analysis on growth rate modeling~\cite{hoffmann2002relativegrowth, marti2017correlations}, we model an organization $v$'s relative growth at year $\timestamp{i}$ as the difference between the logarithms of $v$'s PhD influx and outflux at year $\timestamp{i}$.
We normalize this difference by the number of PhDs working at $v$ during year $\timestamp{i}$ before in- or out-transitions:
 \begin{align}
   \relativeGrowth(v, \timestamp{i}) &= \frac{\textrm{log (\# PhDs joining $v$)}-\textrm{log (\# PhDs leaving $v$)}}{\textrm{log (\# PhDs at $v$) + 1}} \nonumber \\
   &=\frac{\textrm{log}( \sum_u \flowVolume{u}{v} + 1) - \textrm{log}( \sum_w \flowVolume{v}{w} + 1)}{\textrm{log}(\flowVolume{v}{v} + 1) + 1}, \nonumber
    % \label{eq:net-hiring}
\end{align}
where smoothing is used to address noise and correct for zeroes. 
$\relativeGrowth$ is oriented toward fast-growing authorities, so we apply this value to  \emph{incoming} edges of nodes in $\transitionNetwork$.
Since we ultimately want a value between 0 and 1 to retain the same edge-weighting scale as $\resources$ and $\retention$, we transform edges with a normalized exponential function of $\relativeGrowth$:
\begin{align}
    \newFlowVolume{u}{v}=\frac{\flowVolume{u}{v} \cdot \textrm{exp}[\gamma\cdot\relativeGrowth(v, \timestamp{i})]}{\textrm{max}_{v, \timestamp{i}}(\textrm{exp}[\gamma\cdot\relativeGrowth(v, \timestamp{i})])}.
    \label{eq:normalized-relative-growth}
\end{align}
In the above formulation, $\gamma$ controls the steepness of the exponential growth curve (Fig.~\ref{fig:relative-growth}).
For our examples and analyses, we weight the curve with $\gamma=1.5$, chosen by cross-validation to slightly boost influx toward  fast-growing organizations without over-valuing them.
The denominator in Eq.~\eqref{eq:normalized-relative-growth}, which ensures that each transformation is between 0 and 1, normalizes by the maximum exponential growth observed during year $\timestamp{i}$.

\begingroup
\setlength{\intextsep}{0pt}
\setlength{\columnsep}{7pt}%
\begin{wrapfigure}{r}{0.18\textwidth}
\vspace{-0.2cm}
	\centering
	\includegraphics[width=.9\linewidth]{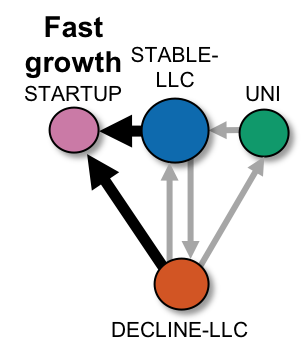}
	\vspace{-.3cm}
	\caption{$\relativeGrowth$ effects.}
	% \vspace{-.5cm}
	\label{fig:relative-growth}
\end{wrapfigure}
\vspace{.1cm}
\noindent \textbf{Example.}
Returning to our example scenario, we  transform $\transitionNetwork$ with the normalized exponential growth function of $\relativeGrowth$.
\nodeD{}'s incoming edge weights are magnified (Fig.~\ref{fig:relative-growth}) and its authority increases from 0.22 to 0.53, overtaking \nodeA{}.
The latter's hub score also increases from 0.13 to 0.43 as several of its employees transition to fast-growing  authority 
\nodeD{}.

\endgroup

\begin{figure}[t]
	\centering
	\includegraphics[width=.7\linewidth]{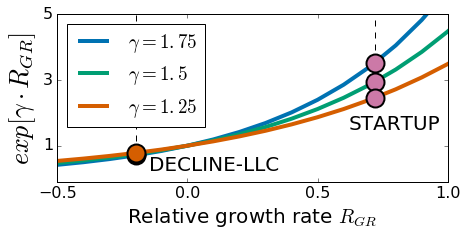}
	\caption{Relative growth of \nodeC{} and \nodeD{}.}
% 	\vspace{-.4cm}
	\label{fig:exp-curve}
\end{figure}

\subsection{Unifying $\resources$, $\retention$, and $\relativeGrowth$}
\label{sec:putting-together}
\begin{figure}[t!]
	\centering
	\includegraphics[width=.9\linewidth,trim={.1cm .1cm .1cm .1cm},clip]{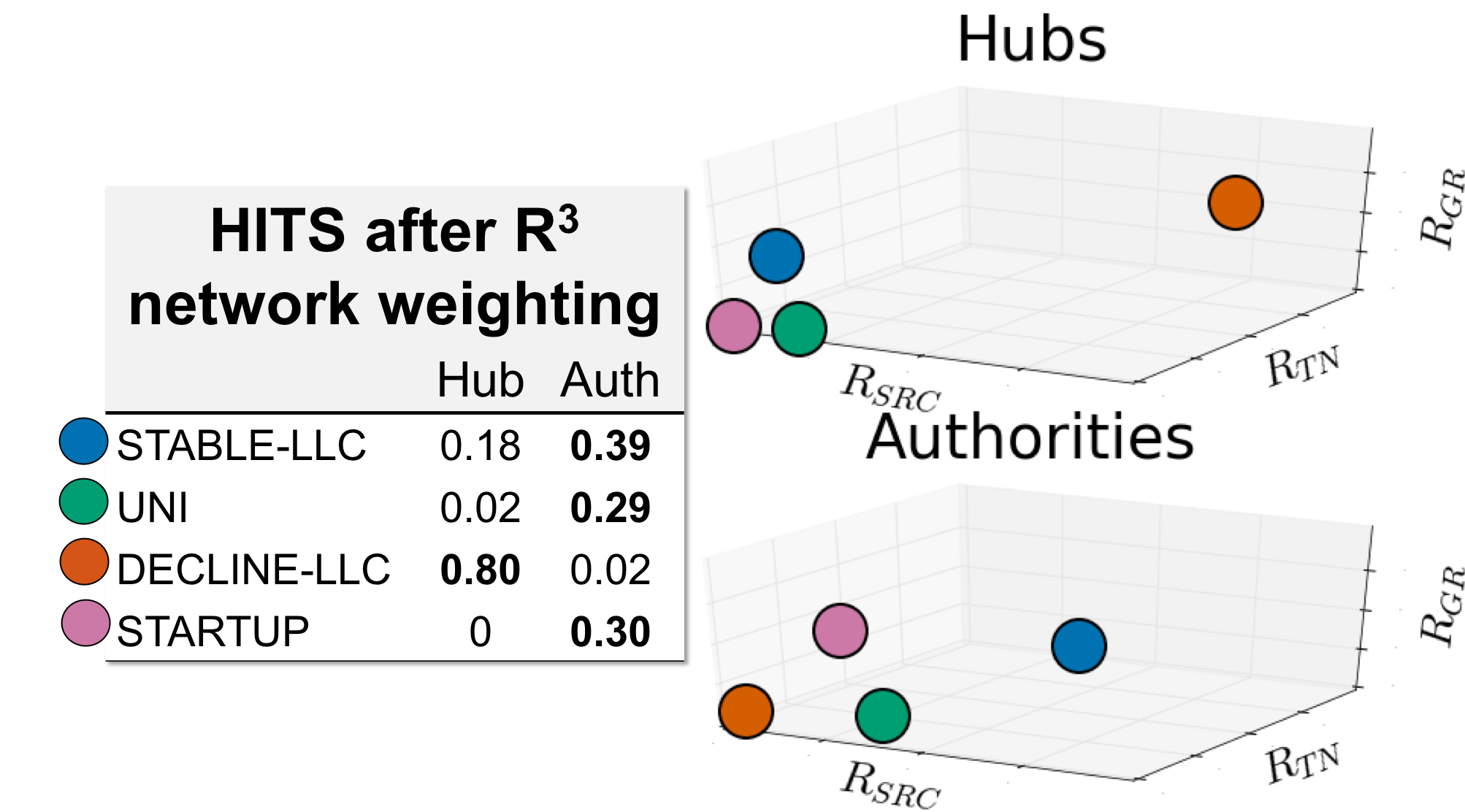}
	\caption{HITS scores from Fig.~\ref{fig:r3-model} placed in the three separate dimensions of \modelName{} (points), and unified (table).
	The hubs and authorities reflect the nuances of the system's dynamics better than in Fig.~\ref{fig:r3-model}.}
 	\vspace{-.2cm}
	\label{fig:r3-scores-3d}
\end{figure}

Thus far we have demonstrated each \emph{R}'s separate effects.
However, ultimately we apply all three \emph{R}'s on the same graph in succession to obtain a \emph{single} set of hubs and authorities rather than different rankings per \emph{R}.
In our running example, the unified \modelName{} scores, which reflect the different temporal career dynamics captured by \modelName{}, are shown in the table of Fig.~\ref{fig:r3-scores-3d}.
Under the unified model, the largest employer, \nodeA{}, no longer dominates the authorities.
Instead, authority is distributed evenly among the three organizations $\nodeA{}$, $\nodeB{}$, and $\nodeD{}$, reflecting the unique importance of each organization in the system.
While $\nodeC{}$ still dominates the hubs, $\nodeA{}$'s hub score also increases because it is a ``stepping stone'' to the fast-growing authority $\nodeD{}$.
% By contrast, $\nodeB{}$'s hub score decreases due to its high retention.
Overall, these changes allow for more meaningful analysis of the system's dynamics because they reflect different kinds of real-world organizational importance.

\section{\modelName{}-driven Analysis}
\label{sec:transitions-trajectories}
\begin{figure*}% [t!]
	\centering
	\includegraphics[width=.99\linewidth]{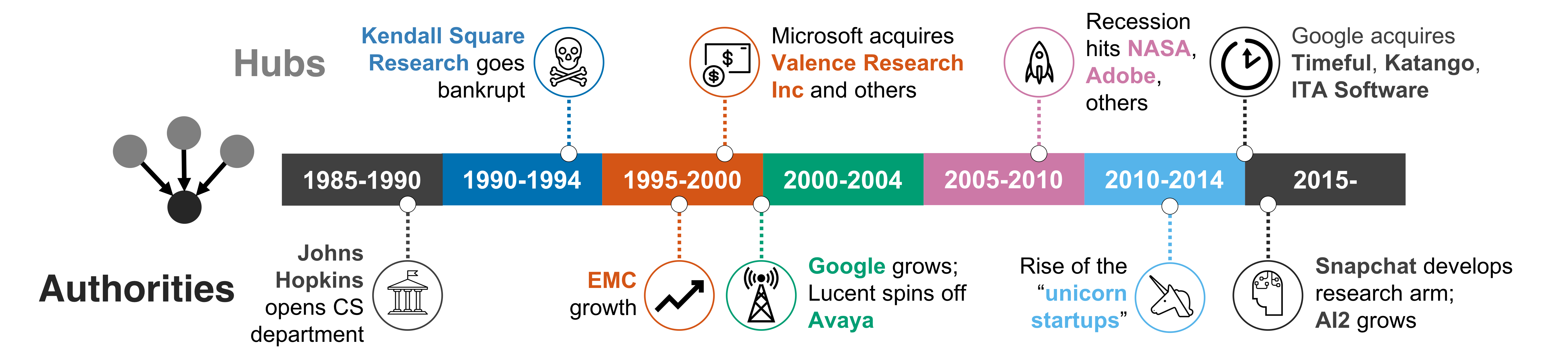}
    \caption{\modelName{} hubs and authorities (bolded) with their respective historical context. 
	The shown organizations' important historical moments were only revealed after using HITS on the \modelName{} transition network as opposed to the aggregate flow network.
	Each of these organizations became a top-50 hub or authority after applying \modelName{} to $\transitionNetwork$. }
% 	\vspace{-.4cm}
	\label{fig:timeline}
\end{figure*}

With \modelName{} defined, we analyze our unique career trajectory dataset and demonstrate \modelName{}'s nuance and versatility.
To do so we: (1) explore the diverse kinds of institutional ``importance'' that \modelName{} captures in computing research; (2) use \modelName{} to characterize significant asymmetry in cross-sector career transitions; and (3) build a strong predictive model with \modelName{} to inform organizational retention. 

\subsection{System-wide evolution}

How do the \emph{R}'s in \modelName{} affect organization rankings on the career transition network $\transitionNetwork$, and what insights can we gain?
Here we %we investigate how the \modelName{} transformations on $\transitionNetwork$ affects HITS rankings, and 
use our model to understand different ways organizations became ``important'' throughout computing research history. 

\vspace{.1cm}
\noindent \textbf{Methodology.} In five-year intervals from 1980 to 2015, we obtain HITS hub and authority rankings for all nodes on the $\transitionNetwork$ and \modelName{} career transition networks. 
% To address the fact that PhD universities are naturally the biggest hubs in our network (all PhDs in our dataset came from one of 50 schools) we create a ``dummy node'' that encompasses all PhD universities, with only outgoing edges to first employers.
% As our goal is to understand the \emph{post}-PhD professional network, we remove this dummy node before computing normalized hub scores (other studies that perform more in-depth analyses of PhD schools and post-PhD academic hiring are reviewed briefly in Sec.~\ref{sec:related}.)
% For the $\resources$ score, we give all new graduates a ``career length'' value of 1 year.
% We then use these scores to compile ``before \modelName{}'' and ``after \modelName{}'' hub/authority rankings for each time interval,
% assigning the same ranks to organizations with the same scores (i.e., more than one organization can have the same ranking, and all organizations with scores of 0 receive the last-ranked place).
We regress the \modelName{} rankings against the $\transitionNetwork$ rankings and identify the nodes with the highest standard error.
These nodes adhere least to the hypothesis that \modelName{} transformations lead to identical HITS scores as $\transitionNetwork$.
They thus capture the most \modelName{}-specific information, their updated rankings a result of the dynamics captured by \modelName{}.
Figure~\ref{fig:timeline} displays several ``important'' historical moments that only became visible after applying the \modelName{} model.
In these moments, the shown organizations' rankings changed significantly with the \modelName{} transformations.
For brevity, here we only cover a small selection of organizations, each of which was outside the top-50 hubs or authorities using HITS on $\transitionNetwork$.
After applying \modelName{}, each of these organizations moved up at least 10 ranks and subsequently ranked as a top-50 hub or authority.
% That said, \modelName{} also moves several organizations \emph{down} in rankings, but we do not cover them here.

% \new{However, we choose the smoothing parameter $\alpha$ such that we avoid making the sigmoid too steep, which in this case would overly penalizing new graduates and heavily weight those who have worked a long time. In our analyses, we set $\alpha=\overline{\ell(\timestamp{i})}/{2}$.}

\vspace{0.1cm}
\noindent \textbf{Results.} 
\noindent (1) \textit{Hubs.} 
The \modelName{} network exposes a variety of organizations, some well-known and some relatively obscure, that were important hubs for different reasons.
The hub timeline in Fig.~\ref{fig:timeline} first shows the supercomputer company Kendall Square Research (KSR).
Founded in the late 1980s and bankrupt by 1994, KSR's short existence and subsequent low retention ($\retention$) contributes to its increased hub ranking, moving from the 96th-ranked hub on $\transitionNetwork$ to the 38th on \modelName{}.
A few years later, \modelName{} designates a host of small companies---Valence Research Inc, Softway Systems, VXtreme Inc, and Vermeer Technologies---as top-20 hubs.
The common theme here is that they all existed for a short time (in consequence, low $\retention$) before acquisition by Microsoft, their employees thus transitioning to one of the top authorities of the era.

In the mid to late 2000s, we observe an increased out-flow of experienced researchers ($\resources$) from several government and industry organizations.
Adobe, VMWare, Disney Research, and NASA become top-20 hubs, which may be in part related to the global economic crisis.
According to Wikipedia, in 2008 Adobe laid of 8\% of its workforce and VMWare fired its CEO after disappointing financial performance.
NASA's funding cuts in the 2000s are also well-documented~\cite{ferguson2013nasa}.
The most recent discovered hubs shown are, like the startups acquired by Microsoft earlier, small short-lived companies that were led by CS PhDs and bought by large companies.
Figure~\ref{fig:timeline} shows Katango and Timeful, two recent acquisitions by Google that moved from rank 100+ to top-30 \modelName{} hubs.
Table~\ref{table:top-n-rankings} highlights ITA Software, also acquired recently by Google.

\vspace{.1cm}
\noindent (2) \textit{Authorities.}
The HITS authority rankings on \modelName{} also lead to interesting discoveries not captured by $\transitionNetwork$.
In the late 80s, \modelName{} designates Johns Hopkins University as the first-ranked authority, which is unusual because universities naturally have lower in-flow.
We found that computer science at Johns Hopkins  officially became a department in 1986, whereas before computing had been part of statistics and operations research. % \footnote{\url{https://www.cs.jhu.edu/30years/2016/08/09/history-snap-shot/}}.
This naturally led to an increased growth rate ($\relativeGrowth$) and an influx of academics ($\resources$) to a school that, prior to 1987, had no representation in our dataset.

In the 90s, EMC (later bought by Dell) becomes a top-20 authority around the time of its rapid growth to billions of dollars in revenue.
This is reflected by its sudden large representation in our dataset in 1997 ($\relativeGrowth$).
\modelName{} next captures the $\resources$ flow to Avaya via its spinoff from the telecommunications giant Lucent in 2000, as well as Google's rise.
Indeed,  \modelName{} ``discovers'' Google before its mid-00s representation increase in our dataset: its authority ranking moves from 45th on $\transitionNetwork$ to 4th on \modelName{} in the late 90s.
In the following decade, \modelName{} rankings differ even further from $\transitionNetwork$ as experienced PhDs ($\resources$) moved from established large hubs like Microsoft, Google, and IBM to small, fast-growing ``unicorn'' startups ($\relativeGrowth$).
Twitter, Dropbox, Snapchat, Square, and Uber all move from rank 100+ to the top-50 in or after the late 00s.
Others of the same vein include Baidu, Netflix, Light, Magic Leap, Databricks, and the nonprofit Allen Institute for Artificial Intelligence (AI2).

\vspace{.1cm}
\noindent (3) \textit{Top-10 rankings.} 
Beyond observing the organizations for which rankings changed the most, we report the top hubs and authorities for two 10-year segments in Table~\ref{table:top-n-rankings}.
The bolded names appear only in the \modelName{} top 10.
We note the mix of organizations across sectors and sizes---some that persisted or rose in rankings, some that dropped or disappeared altogether, some that moved from top-ranked authorities to top-ranked hubs---throughout the years.
\modelName{} is thus capable of capturing various meanings of organizational  ``importance''  despite differences in sector, size, and hiring volume.
That said, %we conclude here by noting that 
while \modelName{} shows that different organizations can gain importance in their own rights, these rankings are of course subjective.
This is especially true in an opportunity-rich field where  ``importance'' is highly dependent on individual goals and interests.

\begin{table}[h!]
    \centering
    \caption{Top employer hub and authority rankings for two time windows.
    The names in bold are those that only appear in their respective top 10s in the \modelName{} transition network.}
    % \vspace{-.8cm}
    \label{table:top-n-rankings}
    \def\arraystretch{0.9}
    \begin{tabular}{ c c  c c  c  }
    \toprule & \multicolumn{2}{c }{\bf\large 1995-2005} & \multicolumn{2}{c}{\bf\large 2005---2015} \\
    \cmidrule(lr){2-3}
    \cmidrule(lr){4-5}
          & {\bf Hub} & {\bf Auth.} & {\bf Hub} & {\bf Auth.} \\
          \toprule
          1 & IBM & IBM & Microsoft & Google \\
          2 & Microsoft & Intel & Yahoo & Microsoft \\
          3 & Intel & Google & Intel & IBM \\
          4 & Bell Labs & Microsoft & Google & Intel \\ % \midrule
          5 & Compaq & Siemens & Siemens & Facebook \\ 
          6 & Sun & {\bf EA} & {\bf ITA Software} & Yahoo \\
          7 & HP & CMU & Amazon & {\bf Snapchat} \\
          8 & {\bf NASA} & MIT & {\bf Apple} & Qualcomm \\
          9 & {\bf Disney} & {\bf Ask.com} & IBM & {\bf MIT} \\ 
          10 & {\bf Docomo} & HP & HP & Amazon \\
          \bottomrule
    \end{tabular}
\end{table}

\subsection{Cross-sector career movement}
\label{sec:past-present}

%  \emph{organization-level} to \emph{individual-level} analysis.
% To this end we use \modelName{} to characterize aspects of individual computing research career \textbf{trajectories}, which appear as 
% ``trails'' or ordered sequences of organizations $(v_1^{\timestamp{1}}, \hdots, v_n^{\timestamp{n}})$ in the transition network. 
% We turn here % from observing the evolution of an entire professional system, anchored by a few key organizations, 
We next study career movement across sectors.
In investigating what \modelName{} reveals about the ``prestige'' associated with cross-sector transitions, we discover insights into \emph{how} people transition between employment sectors and \emph{when} they make these transitions.

\vspace{.1cm}
\noindent \textbf{Methodology.}
Given a year $t$ and an organization $v$, we obtain $v$'s \modelName{} HITS rankings for the five-year interval preceding $t$.
We do this to obtain $v$'s most relevant rankings.
As we showed in the previous section, institutional importance in computing changes quickly, and rankings from more than a few years ago may not be relevant. 
Moreover, due to relatively few transitions per year, a yearly granularity is not appropriate for our analysis.

Note in this section that we consider several thousand organizations in ranking, unlike most well-known ranking systems (i.e., university rankings) that only consider a few hundred institutions.
In accordance with the size and complexity of the computing research professional system, \modelName{} ranking differences between organizations naturally vary more than ranking differences between organizations in smaller-scale studies.

% {Finally, although postdoc records in our dataset are few,
% % and we do not empirically find that including or excluding postdocs from any of our analyses significantly changes our results, we choose to exclude postdocs in this section. 
% we exclude them here to avoid lower rankings for postdoc-granting institutions because postdoc positions lead to lower organizational resource ($\resources{}$) and retention ($\retention{}$) scores.}
% % Postdocs are unique temporary positions usually held in academia right after PhD graduation.}

\vspace{.1cm}
\noindent \textbf{Results.}
% \emph{(1) Inter-sector movement.}
A common narrative in computing research careers is that it is easier to transition to industry from other sectors---in particular, from academia---than vice-versa due to differences in factors like salary and work-life balance.
Without making any claims about the causes of this phenomenon, we do find that our data uphold the narrative.
Out of all cross-sector career transitions in our dataset (16.3\% of all transitions), nearly two-thirds are to industry.

That said, PhDs transitioning \emph{from} industry appear to gravitate toward more ``prestigious'' institutions than their current industry employers.
Without transforming $\transitionNetwork$ with \modelName{}, a PhD transitioning from industry to academia or government moves up on average 34 ranks in HITS authority rankings, whereas a PhD transitioning to industry from academia or government moves \emph{down} on average 47 ranks (we exclude postdocs from these analyses).
Indeed, around 15\% of all PhDs leaving industry in our dataset go to the highly-ranked Stanford, UC Berkeley, MIT, and Carnegie Mellon.
Moreover, PhDs transitioning \emph{to} industry often transition from top-ranked schools to startups, which are naturally ranked lower.
Even those moving up in rankings do not usually move up significantly because of their school's already-high rankings.

\begin{figure}[t!]
	\centering
	\includegraphics[width=.98\linewidth, trim={.1cm 0.2cm .1cm 1cm},clip]{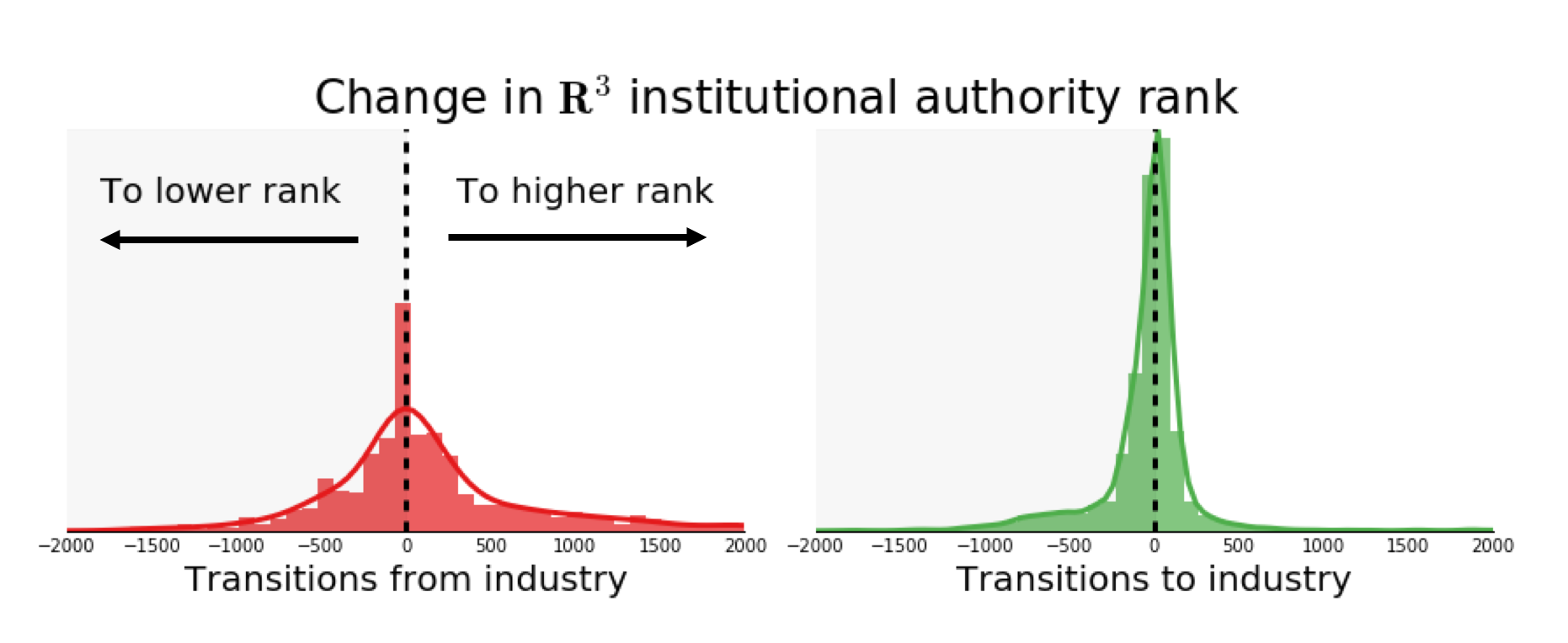}
	\caption{Transitions \emph{from} industry are usually made to higher-ranked organizations.}
	\vspace{-.4cm}
	\label{fig:rank-change-academia-industry}
\end{figure}

The \modelName{} transformations magnify these differences.
A career transition from industry in the \modelName{} network results in an average authority ranking gain of 127 places.
This is partially because the resource factor ($\resources$) captures the imbalance of experience level in these transitions.
The average career lengths of PhDs who transition to and from industry---2.6 years versus 8.23 years respectively---are significantly different
($p\ll.00001$,
%$8.79\times10^{-262}$, 
two-sided $t$-test).
This suggests that PhDs in industry tend to establish themselves first before leaving it, whereas those in academia and government  more often move to industry at earlier stages in their careers.

\begin{figure}% {r}{0.25\textwidth}
	\centering
	\includegraphics[width=.75\linewidth]{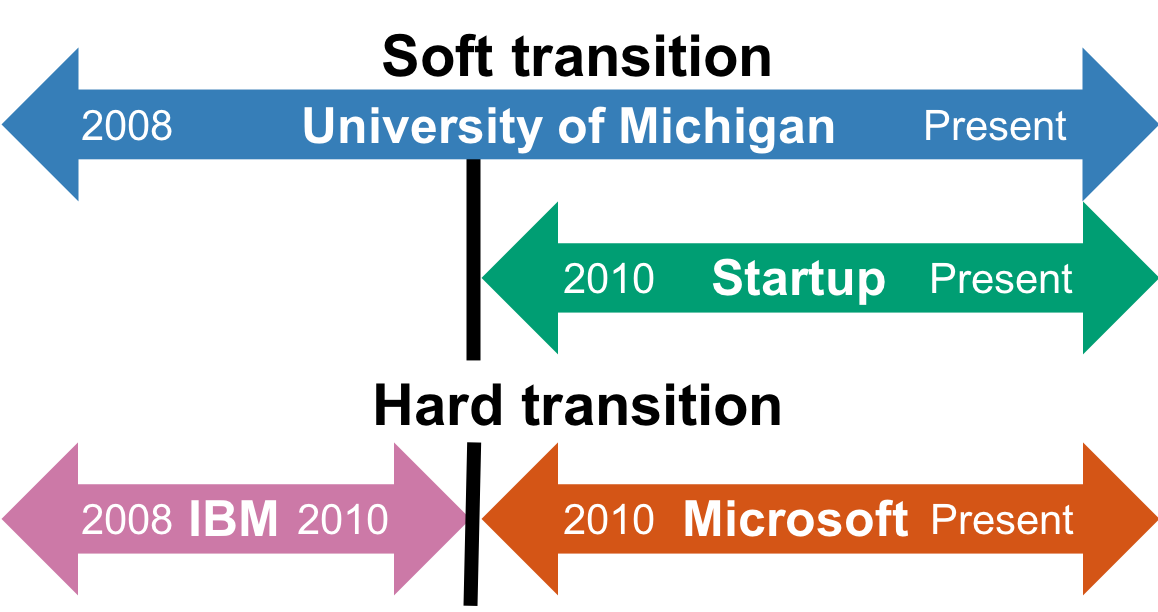}
	\caption{Hard/soft transition examples.}
	% 	\vspace{-.4cm}
	\label{fig:hard-soft-transition-definition}
\end{figure}

\begin{figure}% [h!]
	\centering
	\includegraphics[width=.8\linewidth]{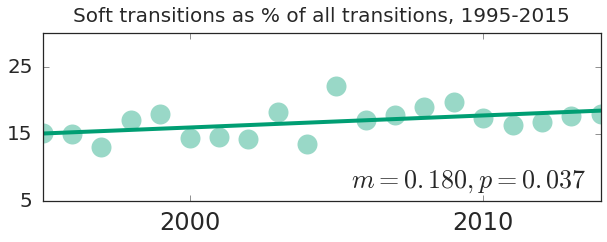}
	\caption{Side projects are becoming more common: percentage of yearly transitions that were ``soft'', 1995 to 2015 (fit line slope $m$=0.180; $p$-value=0.037).}
	% 	\vspace{-.6cm}
	\label{fig:hard-soft-transitions}
\end{figure}

Notably, about 1 in 10 PhDs making a career transition from industry go to consulting or advising, positions often taken up in parallel with other engagements.
Indeed, inter-organization movement is fluid, made complex by the diverse and myriad opportunities in computing for collaboration.
To capture this, we further categorize career transitions into \textbf{hard transitions}, made when an employee \emph{leaves} one organization before joining another, and \textbf{soft transitions}, made when an employee joins an organization \emph{without} leaving her previous employer (Fig.~\ref{fig:hard-soft-transition-definition}).
Soft transitions make up around 21\% of all transitions in our dataset.
Such transitions have been increasing slowly, with some evidence of a linear upward trend  since the mid-90s (Fig.~\ref{fig:hard-soft-transitions}).
This suggests that multiple venues of professional engagement via side projects, collaborations, and startups are becoming more common.


\begin{table*}
	\centering
	\caption{Prediction performance metric averages and standard deviations per value of $n$, highest two per category bolded.}
	% \vspace{-.8cm}
	\label{table:avg-classification}
	\def\arraystretch{.8}
	\begin{tabular}{ clccccc    }
		\toprule
		& \textbf{Features} & $n = 1$ & $n = 2$ & $n = 3$ & $n = 4$ & $n = 5$ \\
		\toprule
		\multirow{4}{*}{\textbf{AUC}} & \egoFeatures{} & $0.625\pm0.00$ & $0.637\pm0.00$ & $0.654\pm0.01$ & $0.644\pm0.01$ & $0.656\pm0.02$ \\
		& \egoFeatures{} + $\transitionNetwork$ & $0.639\pm0.01$ & $0.660\pm0.02$ & $0.666\pm0.02$ & $0.658\pm0.03$ & $0.663\pm0.03$ \\
		& \egoFeatures{} + \modelName{} & $\textbf{0.656}\pm0.01$ & $\textbf{0.675}\pm0.02$ & $\textbf{0.677}\pm0.02$ & $\textbf{0.665}\pm0.02$ & $\textbf{0.670}\pm0.03$ \\
		& \textbf{ALL} & $\textbf{0.649}\pm0.01$ & $\textbf{0.668}\pm0.02$ & $\textbf{0.674}\pm0.02$ & $\textbf{0.664}\pm0.02$ & $\textbf{0.669}\pm0.03$ \\
		\midrule
		\multirow{4}{*}{\textbf{F1}} & \egoFeatures{} & $0.357\pm0.05$ & $0.459\pm0.01$ & $0.536\pm0.01$ & $0.574\pm0.01$ & $\textbf{0.601}\pm0.04$ \\ 
		& \egoFeatures{} + $\transitionNetwork$ & $0.396\pm0.00$  & $0.473\pm0.01$ & $0.542\pm0.00$ & $\textbf{0.577}\pm0.01$ & $\textbf{0.601}\pm0.04$   \\ 
		& \egoFeatures{} + \modelName{} & $\textbf{0.404}\pm0.01$ & $\textbf{0.488}\pm0.01$ & $\textbf{0.549}\pm0.01$ & $0.576\pm0.01$ & $0.595\pm0.03$  \\ 
		& \textbf{ALL} & $\textbf{0.398}\pm0.01$  & $\textbf{0.488}\pm0.01$ & $\textbf{0.550}\pm0.00$ & $\textbf{0.578}\pm0.01$ & $\textbf{0.610}\pm0.03$  \\ \bottomrule
	\end{tabular}
\end{table*}

\begin{figure*}% [t!]
	\centering
	\includegraphics[width=\linewidth]{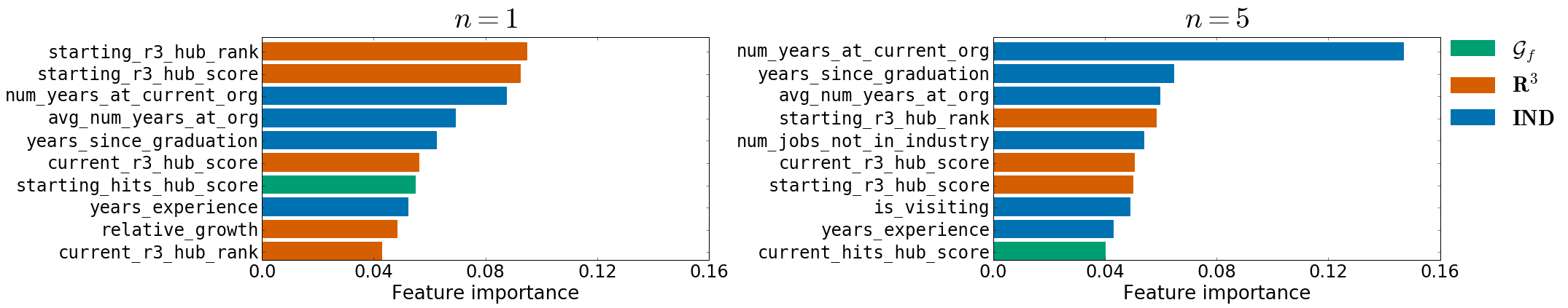}
	% 	\vspace{-.4cm}
	\caption{Feature importances across all features (\textbf{ALL}) for employee transition prediction with $n=1$ and $n=5$.}
	% \vspace{-.4cm}
	\label{fig:feature-importances}
\end{figure*}

With these definitions, we revisit cross-sector career transitions in more detail.
Soft transitions \emph{from} industry are quite common.
Over one-third of all transitions made from industry are soft,
moving \emph{up} on average 64 authority ranks on $\transitionNetwork$ and 174 ranks on \modelName{}.
This suggests that PhDs tend to cross over to prestigious academia and government institutions without fully leaving industry.
By contrast, fewer than 1 in 10 transitions \emph{to} industry are soft, moving \emph{down} on average 133 authority ranks on $\transitionNetwork$ and 248 ranks on \modelName{}.
Nearly half of these rare soft transitions to industry are startup-related, involving a professor taking on a chief role in a budding company.
Two real-world examples are the professor-led startups Timeful and Katango (Fig.~\ref{fig:timeline}), which became top hubs in the 2010s.
% \reminder{bleh} 
% While individuals rarely soft transition to industry, often their higher-than-average skills (on average, 7.6 years of experience pre-transition, corresponding to the $\resources$ factor) confer importance to the corresponding industrial organizations and also ``subtract'' importance when they move to other organizations ($\retention$). 
% For example, a professor-led startup, Timeful, was given a relatively high \modelName{} authority score upon its creation, but also high hubness score when its founder subsequently transitioned to Google, the new (authority) parent company as part of an acquisition.

Overall we observe not-insignificant cross-sector movement, including increasingly popular soft transitions, which highlights the field's various opportunities for collaborations and suggests an increasing connectedness between sectors.
We also observe significant asymmetry in the frequency, timing, and rankings of moves to and from industry, magnified by \modelName{}'s emphasis on experience.
While our data alone cannot point to the \emph{cause} of this asymmetry, a possible explanation is that because soft transitions are less risky or permanent, they allow for bigger leaps, both for those entering highly-ranked academic institutions and those leaving academia for high-risk, high-reward ventures like startups. 

\subsection{Individual retention prediction}
\label{sec:transition-prediction}

Finally, we use \emph{individual} career trajectories, which appear as ordered sequences of organizations $(v_1^{\timestamp{1}}, \hdots, v_n^{\timestamp{n}})$ in the transition network,
% So far we have examined the past and present state of post-PhD computing careers.
along with the \modelName{} network dynamics previously discussed, to predict future employee transitions.
Here we show that \modelName{} boosts the performance of a model with strong predictive power, adding important and interpretable features that can inform organizations seeking to recruit or retain computing PhDs.

\vspace{.1cm}
\noindent \textbf{Methodology.} 
Given a year $t$, can we predict which PhDs $p$ will make a career transition within the next $n$ years?
We group our prediction features (40 total) into 3 categories:
\begin{itemize}
    \item \egoFeatures{}: individual features about $p$'s career trajectory. These include: number of years since graduation; career length in years; number of employers total; average number of years spent with an employer; total years at current employer; number of jobs held in/outside of industry; number of inter-sector transitions; the sectors of $p$'s first and most current employers; number of hard transitions; number of soft transitions; number of postdocs done; and whether $p$'s most current job is senior-level, founder or CEO, professor, researcher, engineer, and/or visiting.
    \item $\transitionNetwork$: features from the aggregate flow career transition network. These include: the HITS rankings and scores of $p$'s current employer $v$, taken from transitions in the 5-year interval up to and including $t$; and the HITS rankings and scores of $v$, taken from transitions in the 5-year interval up to and including the year that $p$ started working at $v$.
    \item \modelName{}: features from our proposed career transition model. These include: the resources ($\resources$) score of $p$ at year $t$; the retention ($\retention$) score of $p$'s current employer $v$ at year $t$;  $v$'s relative growth ($\relativeGrowth$) score during year $t$; and the HITS rankings and scores as with the $\transitionNetwork$ features, but computed on the  \modelName{} network.
\end{itemize}

For those who currently hold more than one job (i.e., soft transitioned to a second job while holding a first), we consider the job started earlier as the ``most current'', since usually soft transitions and side projects occur after full-time employment has already begun. 
% We predict for years $t \in (2000, 2010)$, including only those who graduated at least five years prior to year $t$ and have sufficient career data during year $t$.
As we are interested in predicting immediate career transitions in the modern-day professional system, we vary $n \in [1, 5]$, predicting whether each person will transition between 1 and 5 years ``in the future'', and predict for years $t \in (2000, 2010)$. % including only those who graduated at least five years before $t$ and have sufficient career data during year $t$.

For our prediction tasks, the average ratio of positive labels per value of $n$ is 23.5\%, 32\%, 38\%, 43\%, and 46.5\% for $n\in[1, 5]$ respectively.
For all tasks, we train a gradient boosting tree classifier from the open-source XGBoost
% \footnote{\url{https://github.com/dmlc/xgboost}} 
library, performing a grid search over the %\texttt{eta}, \texttt{subsample}, and \texttt{scale\_pos\_weight}
learning rate (\texttt{eta}), training data subsampling (\texttt{subsample}), and label balance (\texttt{scale\_pos\_weight}) parameters to handle the label imbalance for smaller values of $n$.
We report performance metrics averaged over 10-fold cross-validation in Table~\ref{table:avg-classification} for the \egoFeatures{} features alone, the \egoFeatures{} and $\transitionNetwork$ features together, the \egoFeatures{} and \modelName{} features together, and all groups of features (\textbf{ALL}).

\vspace{.1cm}
\noindent \textbf{Results.} 
The results make it clear that system-wide network dynamics substantially boost prediction performance, justifying our initial choice of using HITS on a career transition network.
Moreover, \modelName{} adds extra power to the model.
The top feature groups by performance are \egoFeatures{} + \modelName{} and \textbf{ALL}, which  do about as well each other.
While AUC remains relatively stable, the higher class imbalance makes the prediction tasks more difficult for smaller values of $n$, highlighting \modelName{}'s strength in improving F1.

Using XGBoost's built-in feature importance tools, we also found that \modelName{} features were consistently considered ``important'' for prediction, and especially so for lower values of $n$ compared to the other feature groups.
% Such features can help employers pinpoint informative retention factors.
The top 10 most important features across all feature groups for $n=1$ and $n=5$ years are given in Fig.~\ref{fig:feature-importances}.
We immediately observe that all of the most important features beyond \egoFeatures{} %, with the exception of \texttt{relative\_growth}, 
are related to an organization's hub scores and ranks.
Most interestingly, we find that for smaller values of $n$, the \modelName{} hub ranks and scores of $p$'s employer \emph{at the time they started working there} are most important, which suggests that employer outflux rate, retention, and volume are about as predictive of short-term retention as individual-level information.

By contrast, for higher values of $n$ the \egoFeatures{} features become more informative.
The most important feature is how many years individuals have worked at their current employers.
This is intuitive given the fast-paced nature of the computing profession and the fact that those who have worked at an employer longer are more likely preparing to transition.
This is especially true in industry, where the employment length mean and variance are shorter.
Indeed, sector-related features become important for higher values of $n$ (\texttt{num\_jobs\_not\_in\_industry}).
Whether or not a person is currently in a visiting position also becomes important, since visiting positions are often limited in duration.

We conclude by noting that while $\transitionNetwork$ features rarely appear in the top 10 most important features, \modelName{} features are consistently important, and $\relativeGrowth$ is one of the top features for $n=1$.
These results show the immediate utility of \modelName{} for retention prediction, which becomes important from an organizational perspective as the demand for computational expertise increases.

\section{Related work}
\label{sec:related}
\vspace{.1cm}
\noindent \textbf{Career path mining.}
Mining professional career paths has attracted recent interest.
One of the first studies to mine career trajectories proposes a similarity measure between professional profiles using temporal sequence alignment on user career paths~\cite{xu2014simcareers}. 
More recently,
~\citet{xu2016talentcircles} detect ``talent circles'' in job transition networks to find qualified candidates for jobs, and~\citet{kapur2016unirankings} apply PageRank on career transition networks as an intermediary step for ranking and recommending universities.
Unlike~\cite{xu2016talentcircles} and~\cite{kapur2016unirankings}, our goals are not recommendation-oriented. 
% is neither to recommend candidates to recruiters nor career moves to job-seekers.
Moreover, while we are not the first to design a weighted HITS scheme~\cite{deguchi2014hitswtn, tseng2015fraudetector}, we are the first, to the best of our knowledge, to propose one for career trajectory mining.

\vspace{.1cm}
\noindent \textbf{Academic career trajectories.}
Most work in academic career trajectory analysis concerns career movement \emph{within} academia.
For example,~\citet{Clauset2015hierarchy} find that academic prestige correlates with higher productivity and better faculty placement, and
\citet{Deville2014geography} find 
that transitions between academic institutions are influenced by career stage and geographical proximity. 
That said, a few studies on the distribution of PhD graduates between academia and industry exist.
For example, ~\citet{Sauermann2012preferences} find in a survey that most students' career preferences shift from academia toward industry over the course of a PhD, and \citet{Balsmeier2014academe} consider how and why scientists \emph{leave} academia.
By contrast, we are interested in \emph{all} directions of cross-sector career movement (Sec.~\ref{sec:transitions-trajectories}), not just transitions from academia.
Moreover, our study is long-term, following individuals beyond their PhDs, and observational, as  we do not use individually-reported preferences to explain the causes of the phenomena we observe. %\reminder{beyond career trajectories}

\vspace{.1cm}
\noindent \textbf{Publication trajectories in computing.}
Several recent works study computing career ``trajectories'' in terms of \emph{publishing} productivity or \emph{citation} counts.
For example,
~\citet{Way2016narrative} study faculty ``productivity trajectories'' in computer science, providing evidence that publishing trends in the field do not follow the conventional ``early peak and gradual decline'' narrative. 
%~\citet{effendy2017trends} present preliminary analysis of trends and evolution in CS publications using the Microsoft Academic Graph dataset.
Most recently,
~\citet{chakraborty2018universal} study scientific ``success trajectories'' in computer science and physics by analyzing paper citation counts.
Similar to~\cite{Way2016narrative}, they question established notions of scientific success, finding multiple distinct trajectories of successful scientific papers beyond the ``early rise'' trajectories of immediately impactful papers. 
These works come from the larger body of research devoted to study of bibliographic data~\cite{newman2009first, hirsch2005index}, which we do not consider here (see discussion). 

\section{Discussion}
\label{sec:discussion}
Our data-driven study of long-term computer science PhD employment dynamics is the first of its kind.
Naturally, many directions for future work remain.
One such direction is increasing the scope of our study in terms of data.
Many important people in computing research  obtained PhDs outside the US.
Some did not obtain a PhD at all.
Although we only considered PhD graduates from a subset of schools to ensure the accuracy of our data, an ideal dataset would include those who made contributions to computing research regardless of degree or background.

A related direction is that of merging bibliographic data with existing career trajectory data.
This task is challenging for large datasets due to the difficulty of entity resolution across databases, which in our case amounts to matching online professional profiles with Google Scholar or DBLP profiles.
However, such data would address questions never before answered: How do publishing rates compare across sectors? 
Do ``impactful'' authors concentrate in ``important'' institutions? 
Is a person's publishing history predictive of their future career transitions?
% \ts{Mention scoping as a way to ensure accuracy}
% An ideal dataset would include both those with PhDs from schools not in US News' top 50 computer science programs, as well as those who made known contributions to computing research regardless of degree or background.
% Furthermore, considering related degree names like ``Applied Mathematics'' and ``Communication Sciences'' could increase representation from the 1970s and 80s, when some PhD programs by other names were effectively computer science.
% \new{Finally, although we designed \modelName{} using our knowledge of, and assumptions about, of the computing research field in particular, \modelName{} could theoretically be applied to any career dataset. Such applications are out of scope of our paper, but would likely reveal many other sector- or field-specific insights.}

Future studies could also perform further data validation.
One inherent limitation of our study is that little standardized data on post-PhD careers exist,
and the data that \emph{do} exist are hard to verify. 
These concerns are not unique to our study, but they are important. 
% While we manually discarded samples that appeared suspicious, 
A future larger-scale study could consider multiple levels of automatic and manual data validation using online CVs, resumes, surveys, and/or news articles as available.
% In particular, although difficult to obtain on a large scale, survey data could increase our model's accuracy, support more fine-grained or specific prediction tasks, and allow more detailed analyses of specific career trajectory features.

A final interesting direction is comparison of different sub-groups in computing research, for example of continents or countries, those with or without a postdoc, and historically underrepresented groups in computing. 
Examining such group-specific differences could lead to actionable organization- and individual-level insights, although such analyses would require self-reporting of gender, race, country of origin, etc.
Again, this direction reduces to the problem of gathering reliable large-scale data.
We hope this will become easier in the future as the topic we study gains traction.

\section{Conclusion}
\label{sec:conclusion}
In this work we examine the career transitions and trajectories of computer science PhDs on the individual, organizational, and sector levels.
We propose \modelName{}, a versatile model for temporal career network dynamics.
Using the HITS link analysis algorithm in conjunction with our \modelName{} model, we:
\begin{itemize}
    \item Provide new insights into the meaning of institutional ``importance'' in computing research careers; 
    \item Reveal a significant asymmetry, from several perspectives, between post-PhD career moves to and from industry; and 
    \item Demonstrate \modelName{}'s immediate utility in supporting prediction of individual career transitions.
\end{itemize}

% We conclude that these analyses may be repeated in the future when more data, made possible by increasing PhD graduation rates, are available. 
While our study is the first of its kind, we conclude by emphasizing that these analyses may be repeated in the future when more data, made possible by increasing PhD graduation rates, are available.
As computing research continues to grow in importance and worldwide presence, this is certain to happen.

% \enlargethispage{2\baselineskip}

\section*{\large Acknowledgements}
The authors thank Rada Mihalcea, Jiongsheng Cai, Shuo Chen, Chengkai Hu, Cole Hudson, Natalia Jenuwine, Jennings Jin, Yinning Wong, and Jiabin Zhu for their early involvement in this project %during its early stages and 
their valuable feedback and contributions.

This material is based upon work supported by the National Science Foundation under Grant No. IIS 1743088, an Adobe Digital Experience research faculty award, a Google scholarship, the University of Michigan, and the GENI project. Any opinions, findings, and conclusions or recommendations expressed in this material are those of the author(s) and do not necessarily reflect the views of the National Science Foundation or other funding parties. The U.S. Government is authorized to reproduce and distribute reprints for Government purposes notwithstanding any copyright notation here on.
% This material is based upon work supported by Google, NSF, and the GENI project.
% This material is based upon work supported by~\new{grants by Google, NSF, etc???}.
% }

\balance

\bibliographystyle{ACM-Reference-Format}
\bibliography{references} 


\begin{thebibliography}{35}

%%% the Web address.

\ifx \showCODEN    \undefined \def \showCODEN     #1{\unskip}     \fi
\ifx \showDOI      \undefined \def \showDOI       #1{#1}\fi
\ifx \showISBNx    \undefined \def \showISBNx     #1{\unskip}     \fi
\ifx \showISBNxiii \undefined \def \showISBNxiii  #1{\unskip}     \fi
\ifx \showISSN     \undefined \def \showISSN      #1{\unskip}     \fi
\ifx \showLCCN     \undefined \def \showLCCN      #1{\unskip}     \fi
\ifx \shownote     \undefined \def \shownote      #1{#1}          \fi
\ifx \showarticletitle \undefined \def \showarticletitle #1{#1}   \fi
\ifx \showURL      \undefined \def \showURL       {\relax}        \fi
% invisible to TeX
\providecommand\bibfield[2]{#2}
\providecommand\bibinfo[2]{#2}
\providecommand\natexlab[1]{#1}
\providecommand\showeprint[2][]{arXiv:#2}

\bibitem[\protect\citeauthoryear{Balsmeier and Pellens}{Balsmeier and
  Pellens}{2014}]%
        {Balsmeier2014academe}
\bibfield{author}{\bibinfo{person}{Benjamin Balsmeier} {and}
  \bibinfo{person}{Maikel Pellens}.} \bibinfo{year}{2014}\natexlab{}.
\newblock \showarticletitle{Who makes, who breaks: Which scientists stay in
  academe?}
\newblock \bibinfo{journal}{{\em Economics Letters\/}} \bibinfo{volume}{122},
  \bibinfo{number}{2} (\bibinfo{year}{2014}), \bibinfo{pages}{229 -- 232}.
\newblock


\bibitem[\protect\citeauthoryear{Bastedo and Bowman}{Bastedo and
  Bowman}{2010}]%
        {bastedo2010usnews}
\bibfield{author}{\bibinfo{person}{Michael~N. Bastedo} {and}
  \bibinfo{person}{Nicholas~A. Bowman}.} \bibinfo{year}{2010}\natexlab{}.
\newblock \showarticletitle{U.S. News \& World Report College Rankings:
  Modeling Institutional Effects on Organizational Reputation}.
\newblock \bibinfo{journal}{{\em American Journal of Education\/}}
  \bibinfo{volume}{116}, \bibinfo{number}{2} (\bibinfo{year}{2010}),
  \bibinfo{pages}{163--183}.
\newblock


\bibitem[\protect\citeauthoryear{Brugere, Gallagher, and Berger-Wolf}{Brugere
  et~al\mbox{.}}{2018}]%
        {BrugereGB18}
\bibfield{author}{\bibinfo{person}{Ivan Brugere}, \bibinfo{person}{Brian
  Gallagher}, {and} \bibinfo{person}{Tanya~Y. Berger-Wolf}.}
  \bibinfo{year}{2018}\natexlab{}.
\newblock \showarticletitle{Network Structure Inference, A Survey: Motivations,
  Methods, and Applications}.
\newblock \bibinfo{journal}{{\em ACM Comput. Surv.\/}} \bibinfo{volume}{51},
  \bibinfo{number}{2}, Article \bibinfo{articleno}{24} (\bibinfo{date}{April}
  \bibinfo{year}{2018}), \bibinfo{numpages}{39}~pages.
\newblock


\bibitem[\protect\citeauthoryear{Chakraborty and Nandi}{Chakraborty and
  Nandi}{2018}]%
        {chakraborty2018universal}
\bibfield{author}{\bibinfo{person}{Tanmoy Chakraborty} {and}
  \bibinfo{person}{Subrata Nandi}.} \bibinfo{year}{2018}\natexlab{}.
\newblock \showarticletitle{Universal trajectories of scientific success}.
\newblock \bibinfo{journal}{{\em Knowledge and Information Systems\/}}
  \bibinfo{volume}{54}, \bibinfo{number}{2} (\bibinfo{year}{2018}),
  \bibinfo{pages}{487--509}.
\newblock


\bibitem[\protect\citeauthoryear{Clauset, Arbesman, and Larremore}{Clauset
  et~al\mbox{.}}{2015}]%
        {Clauset2015hierarchy}
\bibfield{author}{\bibinfo{person}{Aaron Clauset}, \bibinfo{person}{Samuel
  Arbesman}, {and} \bibinfo{person}{Daniel~B. Larremore}.}
  \bibinfo{year}{2015}\natexlab{}.
\newblock \showarticletitle{Systematic inequality and hierarchy in faculty
  hiring networks}.
\newblock \bibinfo{journal}{{\em Science Advances\/}} \bibinfo{volume}{1},
  \bibinfo{number}{1} (\bibinfo{year}{2015}).
\newblock


\bibitem[\protect\citeauthoryear{Clauset, Shalizi, and Newman}{Clauset
  et~al\mbox{.}}{2009}]%
        {ClausetSN09}
\bibfield{author}{\bibinfo{person}{Aaron Clauset},
  \bibinfo{person}{Cosma~Rohilla Shalizi}, {and} \bibinfo{person}{M.~E.~J.
  Newman}.} \bibinfo{year}{2009}\natexlab{}.
\newblock \showarticletitle{Power-Law Distributions in Empirical Data}.
\newblock \bibinfo{journal}{{\it SIAM Rev.}} \bibinfo{volume}{51},
  \bibinfo{number}{4} (\bibinfo{year}{2009}), \bibinfo{pages}{661--703}.
\newblock


\bibitem[\protect\citeauthoryear{Cross}{Cross}{2016}]%
        {economistmoores}
\bibfield{author}{\bibinfo{person}{Tim Cross}.}
  \bibinfo{year}{2016}\natexlab{}.
\newblock \showarticletitle{After Moore's law}.
\newblock  (\bibinfo{year}{2016}).
\newblock
\showURL{%
\url{http://www.economist.com/technology-quarterly/2016-03-12/after-moores-law}}


\bibitem[\protect\citeauthoryear{Deguchi, Takahashi, Takayasu, and
  Takayasu}{Deguchi et~al\mbox{.}}{2014}]%
        {deguchi2014hitswtn}
\bibfield{author}{\bibinfo{person}{T Deguchi}, \bibinfo{person}{K Takahashi},
  \bibinfo{person}{H Takayasu}, {and} \bibinfo{person}{M. Takayasu}.}
  \bibinfo{year}{2014}\natexlab{}.
\newblock \showarticletitle{Hubs and authorities in the world trade network
  using a weighted HITS algorithm}.
\newblock \bibinfo{journal}{{\em PLoS One\/}} \bibinfo{volume}{9},
  \bibinfo{number}{4} (\bibinfo{year}{2014}).
\newblock


\bibitem[\protect\citeauthoryear{Deville, Wang, Sinatra, Song, Blondel, and
  Barabasi}{Deville et~al\mbox{.}}{2014}]%
        {Deville2014geography}
\bibfield{author}{\bibinfo{person}{Pierre Deville}, \bibinfo{person}{Dashun
  Wang}, \bibinfo{person}{Roberta Sinatra}, \bibinfo{person}{Chaoming Song},
  \bibinfo{person}{Vincent~D Blondel}, {and} \bibinfo{person}{Albert-Laszlo
  Barabasi}.} \bibinfo{year}{2014}\natexlab{}.
\newblock \showarticletitle{Career on the Move: Geography, Stratification, and
  Scientific Impact}.
\newblock \bibinfo{journal}{{\em Scientific Reports\/}}  \bibinfo{volume}{4}
  (\bibinfo{year}{2014}).
\newblock


\bibitem[\protect\citeauthoryear{Faloutsos, Faloutsos, and Faloutsos}{Faloutsos
  et~al\mbox{.}}{1999}]%
        {FaloutsosFF99}
\bibfield{author}{\bibinfo{person}{Michalis Faloutsos}, \bibinfo{person}{Petros
  Faloutsos}, {and} \bibinfo{person}{Christos Faloutsos}.}
  \bibinfo{year}{1999}\natexlab{}.
\newblock \showarticletitle{On Power-law Relationships of the Internet
  Topology}.
\newblock \bibinfo{journal}{{\em SIGCOMM Comput. Commun. Rev.\/}}
  \bibinfo{volume}{29}, \bibinfo{number}{4} (\bibinfo{date}{Aug.}
  \bibinfo{year}{1999}), \bibinfo{pages}{251--262}.
\newblock


\bibitem[\protect\citeauthoryear{Ferguson}{Ferguson}{2013}]%
        {ferguson2013nasa}
\bibfield{author}{\bibinfo{person}{Robert~G Ferguson}.}
  \bibinfo{year}{2013}\natexlab{}.
\newblock \bibinfo{booktitle}{{\em NASA's First A}}.
\newblock \bibinfo{publisher}{The NASA History Series}.
\newblock


\bibitem[\protect\citeauthoryear{Fleiss}{Fleiss}{1971}]%
        {fleiss1971kappa}
\bibfield{author}{\bibinfo{person}{Joseph~L. Fleiss}.}
  \bibinfo{year}{1971}\natexlab{}.
\newblock \showarticletitle{Measuring Nominal Scale Agreement Among Many
  Raters}.
\newblock   \bibinfo{volume}{76} (\bibinfo{date}{11} \bibinfo{year}{1971}),
  \bibinfo{pages}{378--}.
\newblock


\bibitem[\protect\citeauthoryear{Getoor and Diehl}{Getoor and Diehl}{2005}]%
        {getoor2005linkmining}
\bibfield{author}{\bibinfo{person}{Lise Getoor} {and}
  \bibinfo{person}{Christopher~P. Diehl}.} \bibinfo{year}{2005}\natexlab{}.
\newblock \showarticletitle{Link Mining: A Survey}.
\newblock \bibinfo{journal}{{\em SIGKDD Explor. Newsl.\/}} \bibinfo{volume}{7},
  \bibinfo{number}{2} (\bibinfo{date}{Dec.} \bibinfo{year}{2005}),
  \bibinfo{pages}{3--12}.
\newblock


\bibitem[\protect\citeauthoryear{Gleich}{Gleich}{2015}]%
        {Gleich15}
\bibfield{author}{\bibinfo{person}{David~F. Gleich}.}
  \bibinfo{year}{2015}\natexlab{}.
\newblock \showarticletitle{PageRank Beyond the Web}.
\newblock \bibinfo{journal}{{\it SIAM Rev.}} \bibinfo{volume}{57},
  \bibinfo{number}{3} (\bibinfo{year}{2015}), \bibinfo{pages}{321--363}.
\newblock


\bibitem[\protect\citeauthoryear{Hirsch}{Hirsch}{2005}]%
        {hirsch2005index}
\bibfield{author}{\bibinfo{person}{Jorge~E Hirsch}.}
  \bibinfo{year}{2005}\natexlab{}.
\newblock \showarticletitle{An index to quantify an individual's scientific
  research output}.
\newblock \bibinfo{journal}{{\em Proceedings of the National academy of
  Sciences of the United States of America\/}} \bibinfo{volume}{102},
  \bibinfo{number}{46} (\bibinfo{year}{2005}), \bibinfo{pages}{16569}.
\newblock


\bibitem[\protect\citeauthoryear{Hoffmann and Poorter}{Hoffmann and
  Poorter}{2002}]%
        {hoffmann2002relativegrowth}
\bibfield{author}{\bibinfo{person}{William Hoffmann} {and}
  \bibinfo{person}{Hendrik Poorter}.} \bibinfo{year}{2002}\natexlab{}.
\newblock \showarticletitle{Avoiding Bias in Calculations of Relative Growth
  Rate}.
\newblock   \bibinfo{volume}{90} (\bibinfo{year}{2002}),
  \bibinfo{pages}{37--42}.
\newblock


\bibitem[\protect\citeauthoryear{Kapur, Lytkin, Chen, Agarwal, and
  Perisic}{Kapur et~al\mbox{.}}{2016}]%
        {kapur2016unirankings}
\bibfield{author}{\bibinfo{person}{Navneet Kapur}, \bibinfo{person}{Nikita
  Lytkin}, \bibinfo{person}{Bee-Chung Chen}, \bibinfo{person}{Deepak Agarwal},
  {and} \bibinfo{person}{Igor Perisic}.} \bibinfo{year}{2016}\natexlab{}.
\newblock \showarticletitle{Ranking Universities Based on Career Outcomes of
  Graduates}. In \bibinfo{booktitle}{{\em ACM KDD}}. \bibinfo{pages}{137--144}.
\newblock


\bibitem[\protect\citeauthoryear{Kleinberg}{Kleinberg}{1999}]%
        {kleinberg1999hits}
\bibfield{author}{\bibinfo{person}{Jon~M. Kleinberg}.}
  \bibinfo{year}{1999}\natexlab{}.
\newblock \showarticletitle{Authoritative Sources in a Hyperlinked
  Environment}.
\newblock \bibinfo{journal}{{\em J. ACM\/}} \bibinfo{volume}{46},
  \bibinfo{number}{5} (\bibinfo{date}{Sept.} \bibinfo{year}{1999}),
  \bibinfo{pages}{604--632}.
\newblock


\bibitem[\protect\citeauthoryear{Koutra, Bennett, and Horvitz}{Koutra
  et~al\mbox{.}}{2015}]%
        {KoutraBH15}
\bibfield{author}{\bibinfo{person}{Danai Koutra}, \bibinfo{person}{Paul~N.
  Bennett}, {and} \bibinfo{person}{Eric Horvitz}.}
  \bibinfo{year}{2015}\natexlab{}.
\newblock \showarticletitle{Events and Controversies: Influences of a Shocking
  News Event on Information Seeking}. In \bibinfo{booktitle}{{\em WWW}}.
  \bibinfo{publisher}{International World Wide Web Conferences Steering
  Committee}, \bibinfo{pages}{614--624}.
\newblock


\bibitem[\protect\citeauthoryear{Koutra, Dighe, Bhagat, Weinsberg, Ioannidis,
  Faloutsos, and Bolot}{Koutra et~al\mbox{.}}{2017}]%
        {KoutraDBWIFB17}
\bibfield{author}{\bibinfo{person}{Danai Koutra}, \bibinfo{person}{Abhilash
  Dighe}, \bibinfo{person}{Smriti Bhagat}, \bibinfo{person}{Udi Weinsberg},
  \bibinfo{person}{Stratis Ioannidis}, \bibinfo{person}{Christos Faloutsos},
  {and} \bibinfo{person}{Jean Bolot}.} \bibinfo{year}{2017}\natexlab{}.
\newblock \showarticletitle{PNP: Fast Path Ensemble Method for Movie Design}.
  In \bibinfo{booktitle}{{\em KDD}}. \bibinfo{publisher}{ACM},
  \bibinfo{pages}{1527--1536}.
\newblock


\bibitem[\protect\citeauthoryear{Koutra and Faloutsos}{Koutra and
  Faloutsos}{2017}]%
        {KoutraF17}
\bibfield{author}{\bibinfo{person}{Danai Koutra} {and}
  \bibinfo{person}{Christos Faloutsos}.} \bibinfo{year}{2017}\natexlab{}.
\newblock \showarticletitle{Individual and Collective Graph Mining: Principles,
  Algorithms, and Applications}.
\newblock \bibinfo{journal}{{\em Synthesis Lectures on Data Mining and
  Knowledge Discovery\/}} \bibinfo{volume}{9}, \bibinfo{number}{2}
  (\bibinfo{year}{2017}), \bibinfo{pages}{1--206}.
\newblock


\bibitem[\protect\citeauthoryear{Landis and G.~Koch}{Landis and
  G.~Koch}{1977}]%
        {landis1977agreement}
\bibfield{author}{\bibinfo{person}{J Landis} {and} \bibinfo{person}{Gary
  G.~Koch}.} \bibinfo{year}{1977}\natexlab{}.
\newblock \showarticletitle{The Measurement Of Observer Agreement For
  Categorical Data}.
\newblock   \bibinfo{volume}{33} (\bibinfo{date}{04} \bibinfo{year}{1977}),
  \bibinfo{pages}{159--74}.
\newblock


\bibitem[\protect\citeauthoryear{Marti, Nielsen, Binkowski, and Donnat}{Marti
  et~al\mbox{.}}{2017}]%
        {marti2017correlations}
\bibfield{author}{\bibinfo{person}{Gautier Marti}, \bibinfo{person}{Frank
  Nielsen}, \bibinfo{person}{Mikolaj Binkowski}, {and}
  \bibinfo{person}{Philippe Donnat}.} \bibinfo{year}{2017}\natexlab{}.
\newblock \showarticletitle{A review of two decades of correlations,
  hierarchies, networks and clustering in financial markets}.
\newblock  (\bibinfo{date}{03} \bibinfo{year}{2017}).
\newblock


\bibitem[\protect\citeauthoryear{Metz}{Metz}{2017}]%
        {nytimesai}
\bibfield{author}{\bibinfo{person}{Cade Metz}.}
  \bibinfo{year}{2017}\natexlab{}.
\newblock \showarticletitle{Tech Giants Are Paying Huge Salaries for Scarce
  A.I. Talent}.
\newblock  (\bibinfo{year}{2017}).
\newblock
\showURL{%
\url{https://www.nytimes.com/2017/10/22/technology/artificial-intelligence-experts-salaries.html}}


\bibitem[\protect\citeauthoryear{Newman}{Newman}{2009}]%
        {newman2009first}
\bibfield{author}{\bibinfo{person}{Mark~EJ Newman}.}
  \bibinfo{year}{2009}\natexlab{}.
\newblock \showarticletitle{The first-mover advantage in scientific
  publication}.
\newblock \bibinfo{journal}{{\em EPL (Europhysics Letters)\/}}
  \bibinfo{volume}{86}, \bibinfo{number}{6} (\bibinfo{year}{2009}),
  \bibinfo{pages}{68001}.
\newblock


\bibitem[\protect\citeauthoryear{Oentaryo, Ashok, Lim, and Prasetyo}{Oentaryo
  et~al\mbox{.}}{2017}]%
        {oentaryo2017jobhops}
\bibfield{author}{\bibinfo{person}{Richard~J Oentaryo},
  \bibinfo{person}{Xiavier~Jayaraj Ashok}, \bibinfo{person}{Ee-Peng Lim}, {and}
  \bibinfo{person}{Philis~Kokoh Prasetyo}.} \bibinfo{year}{2017}\natexlab{}.
\newblock \showarticletitle{On Analyzing Job Hop Behavior and Talent Flow}. In
  \bibinfo{booktitle}{{\em ICDM Data Science for Human Capital Management
  Workshop}}.
\newblock


\bibitem[\protect\citeauthoryear{Safavi, Sripada, and Koutra}{Safavi
  et~al\mbox{.}}{2017}]%
        {SafaviSK17}
\bibfield{author}{\bibinfo{person}{T. Safavi}, \bibinfo{person}{C. Sripada},
  {and} \bibinfo{person}{D. Koutra}.} \bibinfo{year}{2017}\natexlab{}.
\newblock \showarticletitle{Scalable Hashing-Based Network Discovery}. In
  \bibinfo{booktitle}{{\em ICDM}}. \bibinfo{publisher}{IEEE},
  \bibinfo{pages}{405--414}.
\newblock


\bibitem[\protect\citeauthoryear{Sauermann and Roach}{Sauermann and
  Roach}{2012}]%
        {Sauermann2012preferences}
\bibfield{author}{\bibinfo{person}{Henry Sauermann} {and}
  \bibinfo{person}{Michael Roach}.} \bibinfo{year}{2012}\natexlab{}.
\newblock \showarticletitle{Science PhD Career Preferences: Levels, Changes,
  and Advisor Encouragement}.
\newblock \bibinfo{journal}{{\em PLOS ONE\/}} \bibinfo{volume}{7},
  \bibinfo{number}{5} (\bibinfo{date}{05} \bibinfo{year}{2012}),
  \bibinfo{pages}{1--9}.
\newblock


\bibitem[\protect\citeauthoryear{Tong, Faloutsos, and Pan}{Tong
  et~al\mbox{.}}{2006}]%
        {TongFP06}
\bibfield{author}{\bibinfo{person}{Hanghang Tong}, \bibinfo{person}{Christos
  Faloutsos}, {and} \bibinfo{person}{Jia-Yu Pan}.}
  \bibinfo{year}{2006}\natexlab{}.
\newblock \showarticletitle{Fast Random Walk with Restart and Its
  Applications}. In \bibinfo{booktitle}{{\em ICDM}}. \bibinfo{publisher}{IEEE
  Computer Society}, \bibinfo{pages}{613--622}.
\newblock


\bibitem[\protect\citeauthoryear{Tseng, Ying, Huang, Kao, and Chen}{Tseng
  et~al\mbox{.}}{2015}]%
        {tseng2015fraudetector}
\bibfield{author}{\bibinfo{person}{Vincent~S. Tseng},
  \bibinfo{person}{Jia-Ching Ying}, \bibinfo{person}{Che-Wei Huang},
  \bibinfo{person}{Yimin Kao}, {and} \bibinfo{person}{Kuan-Ta Chen}.}
  \bibinfo{year}{2015}\natexlab{}.
\newblock \showarticletitle{FrauDetector: A Graph-Mining-based Framework for
  Fraudulent Phone Call Detection}. In \bibinfo{booktitle}{{\em ACM KDD}}.
  \bibinfo{pages}{2157--2166}.
\newblock


\bibitem[\protect\citeauthoryear{Way, Morgan, Clauset, and Larremore}{Way
  et~al\mbox{.}}{2017}]%
        {Way2016narrative}
\bibfield{author}{\bibinfo{person}{Samuel~F. Way}, \bibinfo{person}{Allison~C.
  Morgan}, \bibinfo{person}{Aaron Clauset}, {and} \bibinfo{person}{Daniel~B.
  Larremore}.} \bibinfo{year}{2017}\natexlab{}.
\newblock \showarticletitle{The misleading narrative of the canonical faculty
  productivity trajectory}.
\newblock \bibinfo{journal}{{\em PNAS\/}} \bibinfo{volume}{114},
  \bibinfo{number}{44} (\bibinfo{year}{2017}), \bibinfo{pages}{E9216--E9223}.
\newblock


\bibitem[\protect\citeauthoryear{Xu, Yu, Yang, Xiong, and Zhu}{Xu
  et~al\mbox{.}}{2016b}]%
        {xu2016talentcircles}
\bibfield{author}{\bibinfo{person}{Huang Xu}, \bibinfo{person}{Zhiwen Yu},
  \bibinfo{person}{Jingyuan Yang}, \bibinfo{person}{Hui Xiong}, {and}
  \bibinfo{person}{Hengshu Zhu}.} \bibinfo{year}{2016}\natexlab{b}.
\newblock \showarticletitle{Talent Circle Detection in Job Transition
  Networks}. In \bibinfo{booktitle}{{\em ACM KDD}}. \bibinfo{pages}{655--664}.
\newblock


\bibitem[\protect\citeauthoryear{Xu, Wickramarathne, and Chawla}{Xu
  et~al\mbox{.}}{2016a}]%
        {xu2016higherorder}
\bibfield{author}{\bibinfo{person}{Jian Xu}, \bibinfo{person}{Thanuka~L.
  Wickramarathne}, {and} \bibinfo{person}{Nitesh~V. Chawla}.}
  \bibinfo{year}{2016}\natexlab{a}.
\newblock \showarticletitle{Representing higher-order dependencies in
  networks}.
\newblock \bibinfo{journal}{{\em Science Advances\/}} \bibinfo{volume}{2},
  \bibinfo{number}{5} (\bibinfo{year}{2016}).
\newblock


\bibitem[\protect\citeauthoryear{Xu, Li, Gupta, Bugdayci, and Bhasin}{Xu
  et~al\mbox{.}}{2014}]%
        {xu2014simcareers}
\bibfield{author}{\bibinfo{person}{Ye Xu}, \bibinfo{person}{Zang Li},
  \bibinfo{person}{Abhishek Gupta}, \bibinfo{person}{Ahmet Bugdayci}, {and}
  \bibinfo{person}{Anmol Bhasin}.} \bibinfo{year}{2014}\natexlab{}.
\newblock \showarticletitle{Modeling Professional Similarity by mining
  Professional Career Trajectories}. In \bibinfo{booktitle}{{\em WWW}}.
\newblock


\bibitem[\protect\citeauthoryear{Yelle}{Yelle}{1979}]%
        {yelle1979learningcurve}
\bibfield{author}{\bibinfo{person}{Louis~E. Yelle}.}
  \bibinfo{year}{1979}\natexlab{}.
\newblock \showarticletitle{The Learning Curve: Historical Review and
  Comprehensive Survey}.
\newblock \bibinfo{journal}{{\em Decision Sciences\/}} \bibinfo{volume}{10},
  \bibinfo{number}{2} (\bibinfo{year}{1979}), \bibinfo{pages}{302--328}.
\newblock


\end{thebibliography}

% \appendix
% \section{Appendix}
% \label{apdx}
% \input{100_appendix}

\end{document}